\documentclass{nature}
\usepackage[utf8]{inputenc}
\usepackage{graphicx}
\usepackage{graphics}
\graphicspath{ {./Figures/} }
\usepackage{duckuments} %
\usepackage{float}
\usepackage{lineno}
\usepackage{comment}
\usepackage{tabularx}
\usepackage{booktabs}
\usepackage{multirow}
\usepackage{amsmath,amssymb}
\usepackage{bbold}
\usepackage[dvipsnames]{xcolor}
\usepackage{mathtools}
\usepackage{braket}

\usepackage[labeled,resetlabels]{multibib}
\newcites{SM}{Supplementary References}

\usepackage{textcomp,gensymb}
\usepackage{prettyref}
\newrefformat{fig}{Fig.~\ref{#1}}
\newrefformat{exfig}{Extended Data Fig.~\ref{#1}}
\newrefformat{sifig}{SI Fig.~\ref{#1}}

\usepackage{hyperref}
\hypersetup{
    colorlinks=true,
    linkcolor=blue,
    filecolor=magenta,      
    urlcolor=cyan,
    citecolor=red,
    pdftitle={Resolving Spectral Gaps and Many-Body resonances in a Moire Superconductor},
}

\usepackage{caption}
\DeclareCaptionLabelFormat{vertline}{#1 #2 $|$}
\usepackage{enumitem}

\makeatletter \let\saved@includegraphics\includegraphics
\AtBeginDocument{\let\includegraphics\saved@includegraphics} 
\renewenvironment{figure}{\@float{figure}}{\end@float}
\makeatother

\definecolor{orange}{rgb}{1,0.5,0}

\newcommand{\beginSI}{%
        \setcounter{table}{0}
        \setcounter{figure}{0}
        \renewcommand{\figurename}{Supplementary Information Fig.}
        \renewcommand{\tablename}{SI Table}
     }

\DeclareMathAlphabet{\mathcald}{U}{dutchcal}{m}{n}
\SetMathAlphabet{\mathcald}{bold}{U}{dutchcal}{b}{n}
\DeclareMathAlphabet{\mathalt}{U}{dutchcal}{b}{n}

\makeatletter
\newcommand{\dummylabel}[2]{\def\@currentlabel{#2}\label{#1}}
\makeatother

\title{Resolving Intervalley Gaps and Many-Body Resonances in Moir\'e Superconductor}

\author{Hyunjin Kim$^{1,2,3}$, Gautam Rai$^{4}$, Lorenzo Crippa$^{4,11}$, Dumitru Călugăru$^{5}$, Haoyu Hu$^{5}$, Youngjoon Choi$^{6}$, Lingyuan Kong$^{1,2}$, Eli Baum$^{1,2}$, Yiran Zhang$^{1,2,3}$, Ludwig Holleis$^{6}$, Kenji Watanabe$^{7}$, Takashi Taniguchi$^{7}$, Andrea F. Young$^6$, B. Andrei Bernevig$^{5,8,9}$, Roser Valent\'i$^{10}$, Giorgio Sangiovanni$^{11}$, Tim Wehling$^{4,12}$, and Stevan Nadj-Perge$^{1,2\dagger}$}
\begin{document}

\maketitle

\begin{affiliations}

  \item T. J. Watson Laboratory of Applied Physics, California Institute of
  Technology, 1200 East California Boulevard, Pasadena, California 91125, USA
  \item Institute for Quantum Information and Matter, California Institute of
  Technology, Pasadena, California 91125, USA
  \item Department of Physics, California Institute of Technology, Pasadena,
  California 91125, USA
  \item Institute of Theoretical Physics, University of Hamburg, Notkestrasse 9, 22607 Hamburg, Germany
  \item Department of Physics, Princeton University, Princeton, New Jersey 08544, USA
  \item Department of Physics, University of California at Santa Barbara, Santa Barbara, California 93106, USA
  \item National Institute for Materials Science, Namiki 1-1, Tsukuba, Ibaraki 305 0044, Japan
  \item Donostia International Physics Center, P. Manuel de Lardizabal 4, 20018 Donostia-San Sebastian, Spain
  \item IKERBASQUE, Basque Foundation for Science, Bilbao, Spain
  \item Institut für Theoretische Physik, Goethe Universität Frankfurt,
 Max-von-Laue-Strasse 1, 60438 Frankfurt am Main, Germany
 \item Institut für Theoretische Physik und Astrophysik and Würzburg-Dresden Cluster of Excellence ct.qmat,
 Universität Würzburg, 97074 Würzburg, Germany
 \item The Hamburg Centre for Ultrafast Imaging, 22761 Hamburg, Germany
  \item[$^\dagger$] Correspondence: s.nadj-perge@caltech.edu
  
\end{affiliations}
\begin{abstract}

Magic-angle twisted multilayer graphene stands out as a highly tunable 
class of moir\'e materials that exhibit strong electronic correlations and robust superconductivity\cite{parkTunableStronglyCoupled2021, haoElectricFieldTunable2021,  parkRobustSuperconductivityMagicangle2022, zhang_promotion_2022}. However, understanding the relations between the low-temperature superconducting phase and the preceding correlated phases established at higher temperatures remains a challenge. 
Here, we employ scanning tunneling microscopy and spectroscopy to track the formation sequence of correlated 
phases established by the interplay of dynamic correlations, intervalley coherence, 
and superconductivity in magic-angle twisted trilayer graphene (MATTG). We discover the existence of two well-resolved gaps pinned at the Fermi level within the superconducting doping range. While the outer gap, previously associated with pseudogap phase\cite{ohEvidenceUnconventionalSuperconductivity2021,kim_evidence_2022}, persists at high temperatures and magnetic fields, the newly revealed inner gap is more fragile in line with superconductivity MATTG transport experiments\cite{parkTunableStronglyCoupled2021, haoElectricFieldTunable2021,  zhang_promotion_2022}. Andreev reflection spectroscopy taken at the same location confirms a clear trend that closely follows the doping behaviour of the inner gap, and not the outer one. Moreover, spectroscopy taken at nanoscale domain boundaries further corroborates the contrasting behavior of the two gaps, with the inner gap remaining resilient to structural variations, as expected from the finite superconducting coherence length. By comparing our findings with recent topological heavy-fermion 
models\cite{songMagicAngleTwistedBilayer2022a,shiHeavyfermionRepresentationTwisted2022}, we identify that the outer gap originates from the splitting of the Abrikosov-Suhl-Kondo resonance\cite{suhlDispersionTheoryKondo1965,abrikosovElectronScatteringMagnetic1965} due to the breaking of the valley symmetry arising from correlation-driven effects. Our results suggest an intricate but tractable hierarchy of correlated phases in twisted multilayer graphene.

\end{abstract}

Materials hosting strongly interacting electrons often exhibit 
a rich interplay of correlated phases that can occur at different
temperatures. Decoupling these phases could be challenging, both 
theoretically and experimentally, when the corresponding temperature 
scales overlap closely, as in twisted multilayer graphene. 
Initial theoretical efforts, based on Hartree-Fock models, established a foundation 
for understanding of correlated phases in the context of flavor polarization and symmetry breaking.
However, entropy measurements revealed fluctuations in spins and valleys, requiring local moments alongside itinerant charge carriers\cite{saitoIsospinPomeranchukEffect2021,zondinerCascadePhaseTransitions2020b,rozenEntropicEvidencePomeranchuk2021} and more recent experiments reported direct 
signatures of the existence of both heavy and dispersive fermions in these 
systems\cite{merinoEvidenceHeavyFermion2024a,batlle-porroCryoNearFieldPhotovoltageMicroscopy2024,ghoshThermopowerProbesEmergent2025}. These 
findings are nominally described by topological heavy fermion models\cite{calugaruThermoelectricEffectIts2024}.
However, the dynamic correlation effects giving rise to heavy fermion band, 
and their influence on the low-temperature symmetry-broken orders have so far 
been experimentally underexplored.

\begin{figure}[p]
\begin{center}
  \includegraphics[width=16cm]{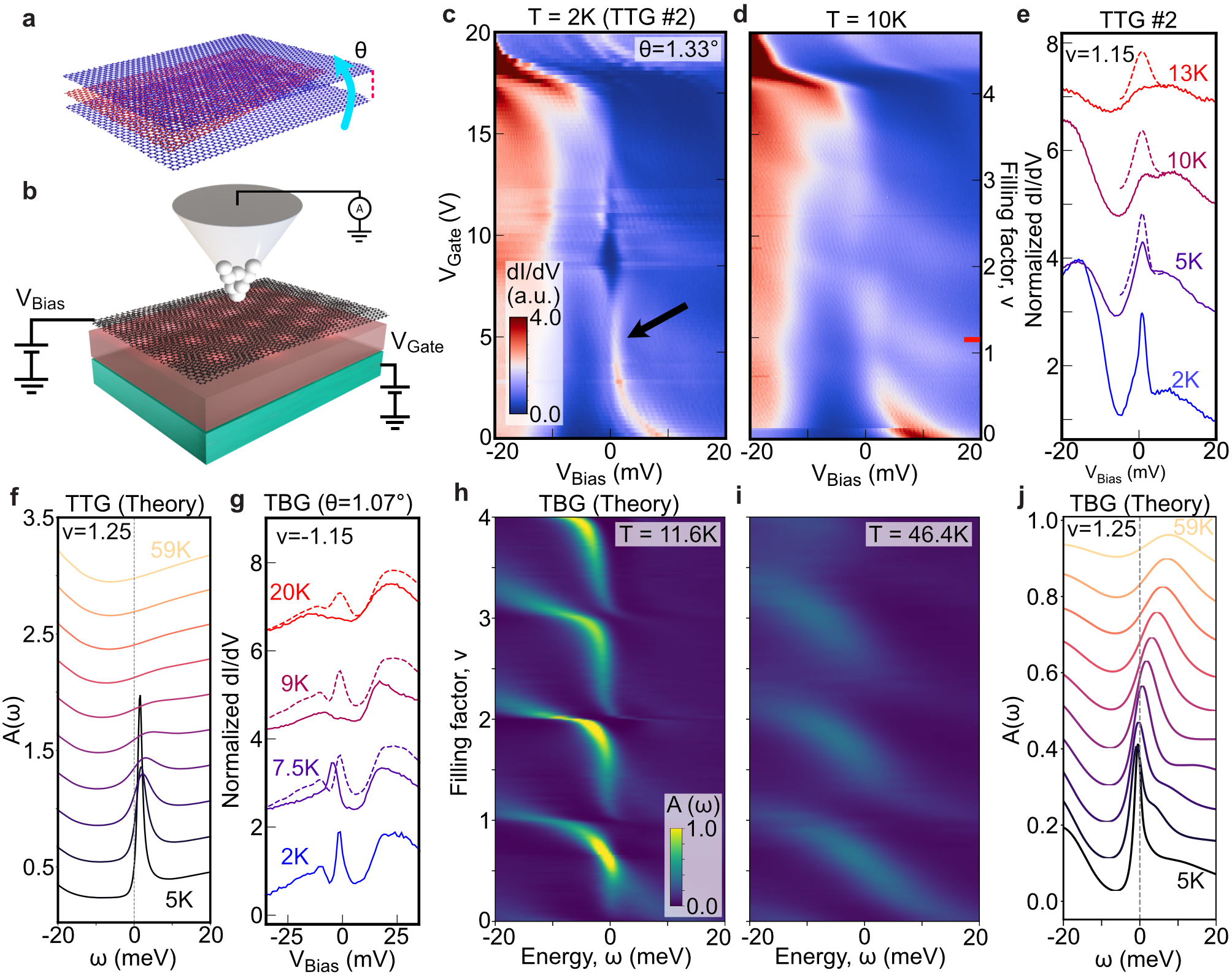}
\end{center}
\caption{ {\bf Onset of many-body Kondo resonance in MATTG.} 
  {\bf a}, Schematic image of alternatively twisted trilayer graphene.
  {\bf b}, Schematic of STM measurement.
  {\bf c}, $V_{\rm Gate}$ dependent $dI/dV$ spectroscopy measured on MATTG device $\#2$, at temperature $T = 2$ K, with twist angle of $\theta = 1.33\degree$. The black arrow indicates the high density of states peak pinned at $E_F$.
  {\bf d}, $V_{\rm Gate}$ dependent $dI/dV$ spectroscopy measured on the same area as \prettyref{fig: fig1}c at a temperature $T = 10$ K. Red line indicates the $V_{\rm Gate}$ where the $dI/dV$ spectra in \prettyref{fig: fig1}e is measured.
  {\bf e}, $dI/dV$ spectra focused on $dI/dV$ peak at  $E_F$ at multiple temperatures. The dashed line is the convolution of the $2$ K spectrum with the Fermi-Dirac function at each temperature.
  {\bf f}, Temperature dependence of the calculated momentum-integrated spectrum function $A(\omega)$ in the symmetric phase of MATTG. Temperature for each curve is $5$K, $11$K, $17$K, $23$K, $29$K, $35$K, $41$K, $47$K, $53$K, $59$K.
  {\bf g}, $dI/dV$ spectra measured on MATBG sample at $\theta = 1.07\degree$ focusing on the $dI/dV$ peak at  $E_F$ at multiple temperatures. The dashed line is the convolution of the $2$ K spectrum with 
 the Fermi-Dirac function at each temperature.
  {\bf h-i}, The DMFT momentum-integrated spectral function $A(\omega)$ in the symmetric phase at low ({\bf g}) and  high ({\bf h}) temperature as a function of electron doping. 
  {\bf j}, Temperature dependence of the calculated momentum-integrated spectrum function $A(\omega)$ in the symmetric phase of MATBG. Temperature for each curve is $5$K, $11$K, $17$K, $23$K, $29$K, $35$K, $41$K, $47$K, $53$K, $59$K.
  }
\label{fig: fig1}
\end{figure}

\prettyref{fig: fig1}a, b show the structure of magic-angle twisted trilayer graphene (MATTG) and the corresponding scanning tunneling microscopy (STM) geometry. 
Due to the mirror symmetry, MATTG band structure can be decomposed into a set of twisted bilayer 
graphene-like flat bands and a dispersive Dirac cone in the absence of a perpendicular electric field\cite{khalafMagicAngleHierarchy2019b,leiMirrorSymmetryBreaking2021,carrUltraheavyUltrarelativisticDirac2020a}. 
Spectroscopic mapping as a function of gate voltage in 
the area with moderate hetero-strain ($\epsilon = 0.2\%$) and twist angle $\theta = 1.33\degree$ (\prettyref{fig: fig1}c) shows an abrupt band structure renormalization occurring at every integer filling factor ($\nu=\pm1,\pm2,\pm3$) of the moir\'e bands. This renormalization is attributed to the formation of the upper Hubbard bands with positive bias as a result of the Coulomb interaction\cite{wongCascadeElectronicTransitions2020a,zondinerCascadePhaseTransitions2020b}.
We focus our attention on the upper van Hove singularity that migrates towards the Fermi level upon electron doping\cite{kerelskyMaximizedElectronInteractions2019b,choiElectronicCorrelationsTwisted2019b,jiangChargeOrderBroken2019a,xieSpectroscopicSignaturesManybody2019b}.
This spectral peak is pinned to $V_{\rm Bias} = 0$ mV, i.e., at the Fermi level for a range of fillings, and overlaid with band renormalization coming from the cascade transition. At elevated temperatures, the density of states still renormalizes at every integer filling, but the corresponding spectral peak is no longer pinned at the Fermi level and, moreover, its height rapidly diminishes (\prettyref{fig: fig1}d,e). 
We note that this behaviour is also observed in magic-angle twisted bilayer graphene (\prettyref{fig: fig1}f) and does not occur near charge neutrality, where peaks in density of states do not show a significant
decrease at elevated temperatures (\prettyref{exfig: Temp}).

The strong temperature dependence of the zero-bias pinned $dI/dV$ peak relative to the upper Hubbard band suggests the presence of correlation effects not captured by
thermal broadening of a simple van Hove singularity.
This zero-bias pinned density of states with a characteristic onset 
temperature 
is reminiscent of an Abrikosov-Suhl-Kondo-type of resonance found in heavy-fermion compounds and
strongly-correlated materials\cite{jiaoMagneticDefectProbes2018,yeeImagingKondoInsulating2013,seiroEvolutionKondoLattice2018,zhangManyBodyResonanceCorrelated2020}.
In this scenario, resonant states pinned at the Fermi energy are a consequence of renormalization effects resulting from dynamical correlations. 
Dynamical mean-field theory (DMFT) calculations based on the topological heavy fermion model (THF) of MATBG\cite{songMagicAngleTwistedBilayer2022a,shiHeavyfermionRepresentationTwisted2022} reproduce well the appearance of the zero-energy resonance for sufficiently low temperatures as well as the neighboring Hubbard peaks (\prettyref{fig: fig1}h,i) confirming this picture. 
More detailed temperature dependence (\prettyref{fig: fig1}j) shows that a temperature increase gradually moves the peak away from  $E_F$, as it is also identified from the experiment (\prettyref{fig: fig1}g).
The observed spectroscopic behavior away from integer filling is further indicative of
intermediate valence physics:
1) The Hubbard peaks are highly asymmetric, suggesting intermediate valence physics\cite{horvaticFinitetemperatureSpectralDensity1987}. 2) Upon lowering the temperature, the quasiparticle peak emerges out of the side peak that is closer to the Fermi level\cite{horvaticFinitetemperatureSpectralDensity1987}.
3) The electron compressibility is high throughout the range of fillings where the pinned peak is observed (see SI and Ref. \citenum{raiDynamicalCorrelationsOrder2024}).
Theoretical modeling taking into account dynamical 
correlations in MATTG\cite{yuMagicangleTwistedSymmetric2023,calugaruThermoelectricEffectIts2024} has also been performed by iterative perturbation theory (IPT) calculations where we reproduce the temperature and filling dependence of the zero-energy resonance (\prettyref{fig: fig1}f).

\begin{figure}[p]
\begin{center}
   \includegraphics[width=16cm]{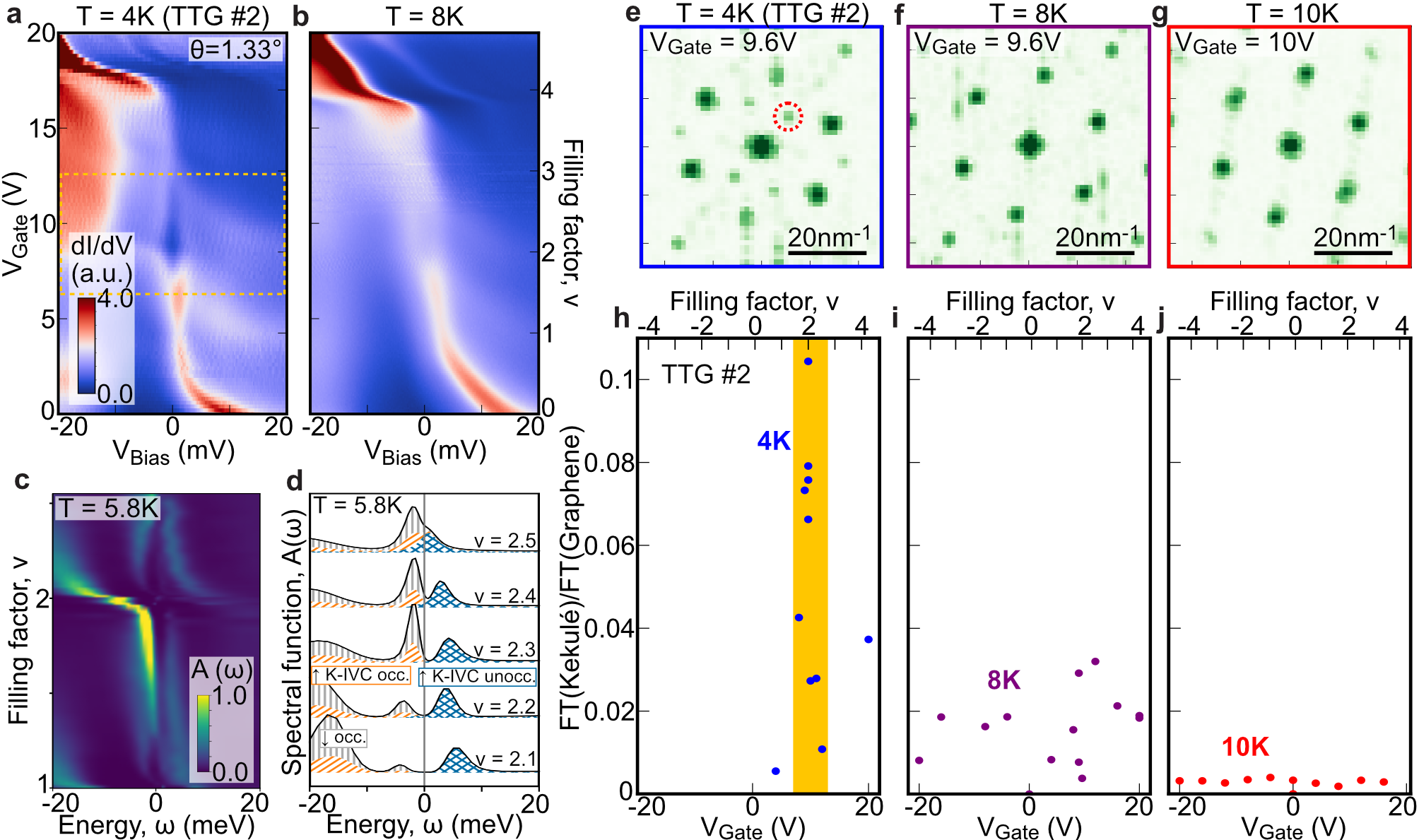}
\end{center}
\caption{
{\bf Onset of inter-valley coherent order in MATTG} {\bf a, b}, $V_{\rm Gate}$ dependent $dI/dV$ spectroscopy measured in MATTG device $\#2$ at temperature $T = 4$ K ({\bf a}), $T = 8$ K ({\bf b}).
{\bf c}, The calculated momentum integrated spectral function $A(\omega)$ in the (K-)IVC ordered phase as a function of doping. The white dashed box emphasizes the gap appearing in similar range of filling where the gap is observed in experiment.
{\bf d}, The calculated momentum-integrated spectral function $A(\omega)$ projected to distinct sectors: grey, orange, and blue shades represent projection to the spin $\downarrow$ state, spin $\uparrow$ K-IVC occupied state, spin $\uparrow$ K-IVC unoccupied state respectively.
{\bf e-g}, Real-space $dI/dV$ map at $\nu = 2.4$ taken at temperature $T = 4$ K ({\bf e}), $T = 6$ K ({\bf f}), $T = 8$ K ({\bf g}). Red dashed circle highlights the FT signal corresponding to Kekul\'e pattern at $T = 4$ K.
{\bf h-j}, Intensity of the Kekulé reciprocal lattice vector normalized by the intensity of the graphene reciprocal lattice
vector extracted at multiple $V_{\rm Gate}$ for $4$ K ({\bf h}), $8$ K ({\bf i}), and $10$ K ({\bf j}). Yellow region highlights the $V_{\rm Gate}$ range where the gap is observed in {\bf a}.
}
\label{fig: fig2}
\end{figure}

We now focus on the MATTG spectrum at a temperature range at which the intervalley-coherent (IVC) ordering onsets (\prettyref{fig: fig2}a,b).  Slightly above the IVC ordering temperature, the quasiparticle peak at zero-bias is smoothly connected in the filling factor range $\nu = 1$--$3$. (\prettyref{fig: fig2}b). On the other hand, below the IVC order temperature, between $\nu = 1.6$--$2.6$, this peak evolves into a gapped spectrum pinned at $E_F$ (\prettyref{fig: fig2}a). The onset of IVC order is established from real-space $dI/dV$ maps followed by Fourier transformation (FT) analysis (\prettyref{fig: fig2}e-g).
Comparing three different temperatures ($4, 8, 10$ K) and multiple fillings (\prettyref{fig: fig2}h-j), we find that IVC order appears, as confirmed by the observation of Kekul\'e reciprocal lattice peaks, at $4$ K between $\nu = 1.6$--$2.6$.  
The one-to-one correspondence of the Kekul\'e distortion and the appearance of the gap at the same filling and temperature suggests a strong connection between the two phenomena (IVC ordering observed in real space and formation of the gap).

Our interpretation of the experimental results agrees with DMFT calculations at low temperatures where inter-valley coherence is stabilized\cite{raiDynamicalCorrelationsOrder2024}.
\prettyref{fig: fig2}c shows the DMFT spectrum of MATBG under Kramers intervalley coherent state (KIVC) for fillings ranging from $\nu = 1 \sim 2.5$. 
We note that the THF model incorporating finite heterostrain\cite{kwanKekuleSpiralOrder2021a, herzog-arbeitmanHeavyFermionsEfficient2025, herzog-arbeitmanKekuleSpiralOrder2025} has been shown to reproduce the gapped spectrum around $\nu = 2$, which bears similarity with the spectrum from KIVC order at zero heterostrain computed here.
The many-body resonance from \prettyref{fig: fig1}j forms a gap around $E_F$,
consistent with our experimental results.
In experiments, we observe the IVC phase only for $\nu = 1.6 \sim 2.6$, coinciding with the filling factor range where the gap spectrum occurs.
\prettyref{fig: fig2}d reveals different flavors dominating the spectral weight below and above the Fermi level. 
Thus, the gap opening mechanism suggested by the DMFT results is analogous to the 
splitting of a Kondo resonance in the Anderson impurity model\cite{bagchiSpinPolarizedScanningTunneling2024,hofstetterGeneralizedNumericalRenormalization2000}, where the zero-bias peak splits as the coherent quasiparticle becomes spin-polarized. 
However, in our case, the splitting is driven by spontaneous IVC ordering rather than an applied magnetic field\cite{bagchiSpinPolarizedScanningTunneling2024,hofstetterGeneralizedNumericalRenormalization2000}.
This intuitive picture suggests that the two $dI/dV$ peaks (for positive and negative energy) are expected to have opposite IVC polarization which is consistent with the reversal of the Kekul\'e pattern in the experiment (\prettyref{exfig: Kek_bias}).
While we focus on one particular twist angle where a strong correlation between gap opening and IVC order is observed, we recall a similar feature has been observed in our previous experiment performed on a different sample\cite{kim_imaging_2023}.

\begin{figure}[p]
\begin{center}
  \includegraphics[width=16cm]{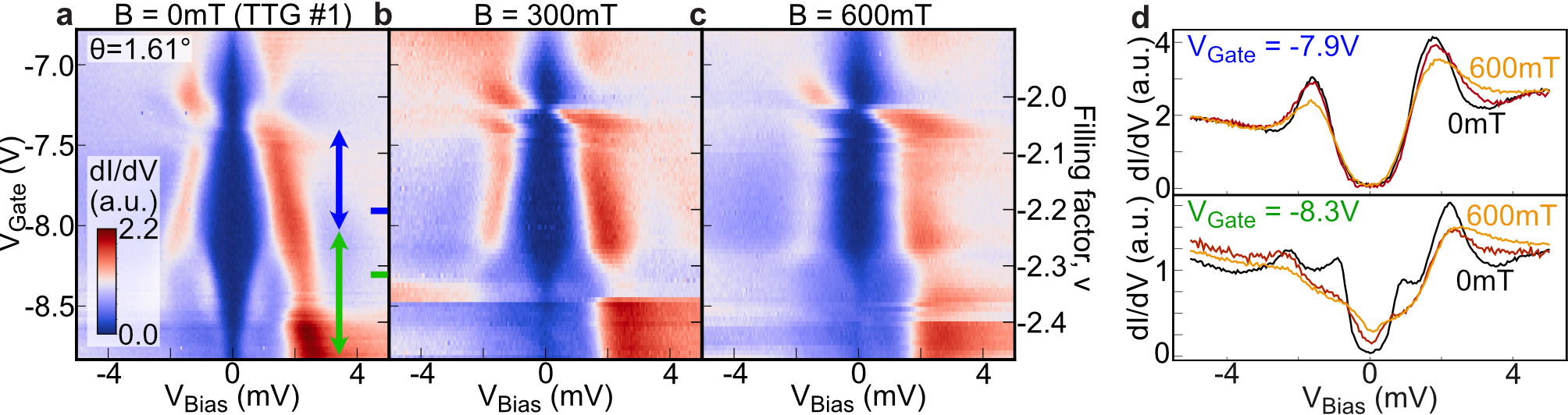}
\end{center}
\caption{{\bf Observation of two gaps pinned to Fermi level in superconducting MATTG.} 
{\bf a-c}, $V_{\rm Gate}$ dependent $dI/dV$ spectroscopy measured in MATTG device $\#1$ ($\theta = 1.59\degree$, $\epsilon = 0.12\%$ area) focused on the correlated gap around $\nu = -2$ under perpendicular magnetic $B = 0$ mT ({\bf a}), $B = 300$ mT ({\bf b}), $B = 600$ mT ({\bf c}).
{\bf d}, Magnetic field dependence of the spectrum (each line corresponding to $B = 0$ mT, $B = 300$ mT, $B = 600$ mT) for the single-gap regime (upper panel) and the two-gap regime (lower panel).
}
\label{fig: fig3}
\end{figure}

\begin{figure}[p]
\begin{center}
   \includegraphics[width=16cm]{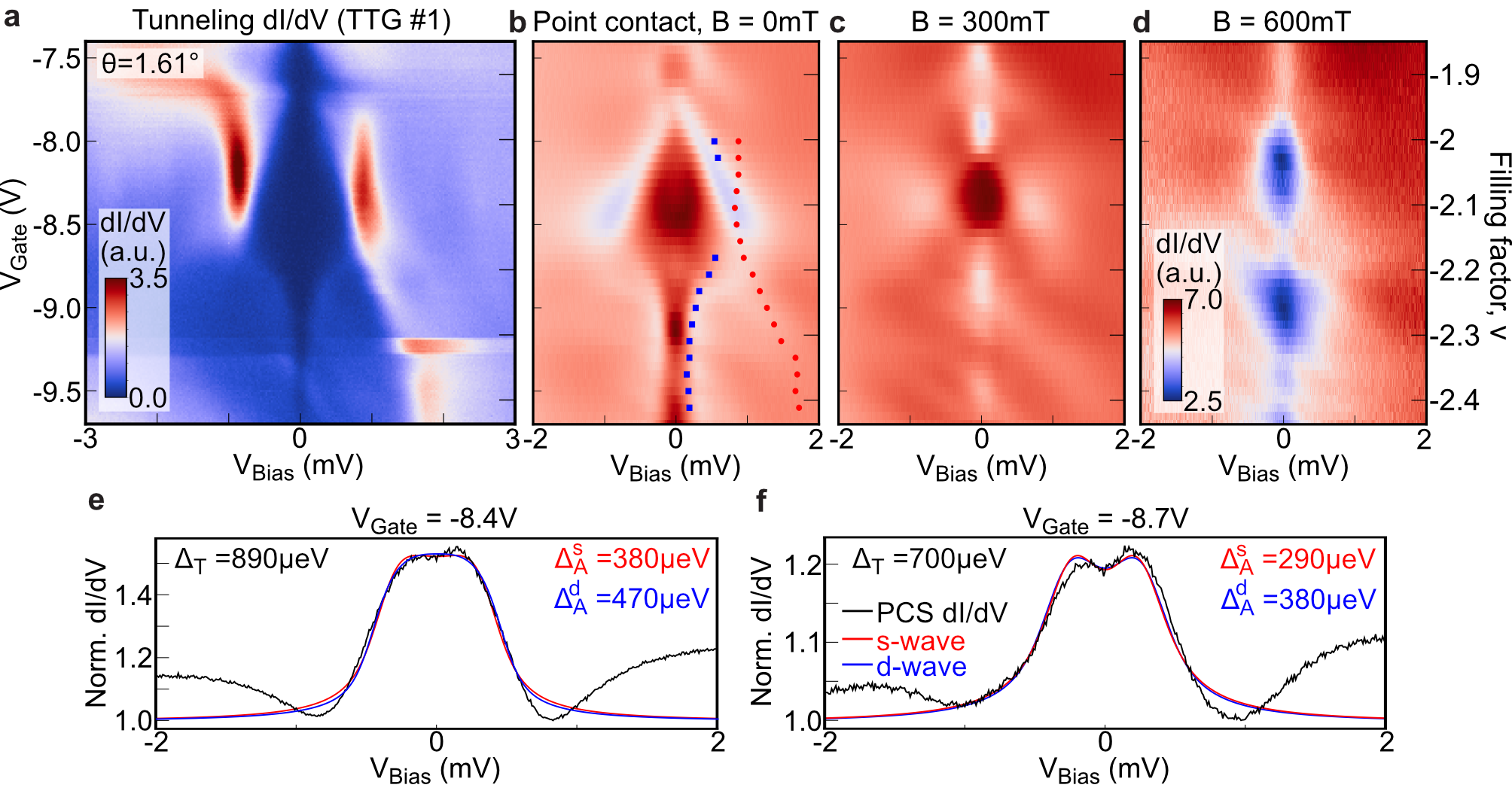}
\end{center}
\caption{{\bf Comparison of tunneling and Andreev signals in the superconductivig phase.} 
{\bf a}, $V_{\rm Gate}$ dependent $dI/dV$ spectroscopy measured in MATTG device $\#1$ ($\theta = 1.61\degree$, $\epsilon = 0.10\%$).
{\bf b-d}, Point contact conductance ($dI/dV$) spectroscopy as a function of $V_{\rm Gate}$ at multiple perpendicular magnetic fields of $B = 0$ mT ({\bf b}), $B = 300$ mT ({\bf c}), $B = 600$ mT ({\bf d}). Blue (red) dots in \prettyref{fig: fig4}b indicate extracted inner (outer) gap size from \prettyref{fig: fig4}a by identifying local maxima in second derivative of $dI/dV$ spectrum.
{\bf e}, Tunneling (black) and point contact (blue) conductance ($dI/dV$) spectrum overlaid for each $V_{\rm Gate} = -8.7$ V, $-8.8$ V, $-8.9$ V, $-9.0$ V.
{\bf e, f}, Point contact conductance ($dI/dV$) taken for single gap regime ({\bf e}), and two gap regime ({\bf f}).
Red and blue curve corresponds to Blonder-Tinkham-Klapwijk formula fit for s-wave (red), d-wave (blue) superconducting order parameter.
}
\label{fig: fig4}
\end{figure}

Upon further lowering the temperature, striking features emerge in the low-energy spectrum.
Here we focus on the $\theta = 1.61\degree$ MATTG sample, where we previously identified an incommensurate Kekulé spiral order between $\nu = -2 \sim -3$
\cite{kim_imaging_2023}. At this twist angle and heterostrain, STM measurements found gaps in the range of filling factors $\nu = -2 \sim -3$ at $2$ K, consistent with the IKS gap formed at the Fermi level.
At $T=400$ mK , we observe additional features developing around the Fermi energy for a filling factor range $\nu = -2 \sim -2.5$ (\prettyref{fig: fig3}a). From $\nu = -2$ to $\nu = -2.25$ (blue arrow in \prettyref{fig: fig3}a), the spectrum shows a single, fully developed gap with $dI/dV$ going to $0$ at $V_{\rm Bias} = 0$ mV. Within this range, the formed gap can be understood 
either to be formed by the IKS state (as discussed above) or by the emergence of superconductivity, as evidenced from the Andreev reflection data in \prettyref{fig: fig4}. Surprisingly, further hole doping results in a splitting of the single coherence peak, indicating the presence of two well-resolved gaps. In the range of hole doping where both gaps are visible (from $\nu = -2.25$ to $\nu = -2.7$, green arrow in \prettyref{fig: fig3}a)
the inner gap shrinks with the hole doping while the outer gap shows a more complex evolution. It first increases until $\nu = -2.4$, where the maximal IVC strength is observed (as measured by the intensity of Kekul\'e FT peaks\cite{kim_imaging_2023}), and decreases with further hole doping until it disappears around  $\nu = -3$. Importantly, the onset temperature of the inner gap at $\nu = -2.3$ is $1.5$ K (\prettyref{exfig: InnerT}), matching the Andreev signal, further confirming its superconducting origin. We note that a similar two-gap feature is observed in TTG device $\#2$ with a different twist angle, showing superconductivity on the electron side (\prettyref{exfig: TTG16}).

The single and double gap regimes also show notable differences with 
applied out-of-plane magnetic fields. In a single gap regime, 
moderately small fields ($B<1$ T) have minimal effect on the gap 
spectrum  (\prettyref{fig: fig3}b-c). In this 
regime, we observe only a slight lowering of the coherence peak height 
and minor changes in gap size, but the system remains fully gapped 
around the Fermi level. This contrasts with the double gap regime, 
where the inner gap is significantly suppressed with the field and 
disappears around $600$ mT. This field value aligns with the critical magnetic field determined through transport measurements in previous MATTG studies\cite{parkTunableStronglyCoupled2021,haoElectricFieldTunable2021,zhang_promotion_2022}.
While the inner gap dramatically changes, the outer gap remains similar in size (with some minor changes in appearance \prettyref{exfig: Switching}).
The resilience of the outer gap under the applied field and the observation of electronic reconstruction in real space at the same doping range and magnetic fields further indicates that the outer gap origin is tied to intervalley-coherent ordering.

To further investigate the nature of the inner gap and corroborate its relation to superconductivity, we focus on point contact spectroscopy measurements\cite{ohEvidenceUnconventionalSuperconductivity2021,kim_evidence_2022} that reveal details about the Andreev reflection process. We note that these measurements have been accompanied by high-resolution tunneling spectroscopy at the same location (\prettyref{fig: fig4}a, see also  \prettyref{exfig: PCS} for details of the point contact spectroscopy measurements).
Gate-dependent contact spectroscopy shows the enhanced conductance around zero bias only for filling factors around $\nu = -3$ to $-2$ (\prettyref{exfig: PCS}) signaling phase coherent transport between the superconducting sample and the normal metallic tip, i.e., Andreev
reflection\cite{deutscherAndreevSaintJamesReflectionsProbe2005}.
Point contact spectra are further analyzed by fitting the Blonder-Tinkham-Klapwijk formula\cite{blonderTransitionMetallicTunneling1982b} in the single and double-gap regime to extract the corresponding gap size $\Delta_{A}$ (\prettyref{fig: fig4}e,f).
The BTK fit shows that both isotropic and nodal order parameters are consistent with the Andreev reflection signal which is difficult to resolve in tunneling spectra due to non-flat normal state density of states.
The behavior of the Andreev signal at finite temperatures (\prettyref{exfig: InnerT}b) and out-of-plane magnetic fields shows its disappearance around $1$K and $600$mT in line with the transport data\cite{parkTunableStronglyCoupled2021, haoElectricFieldTunable2021, zhang_promotion_2022}. 
Importantly, the energy (bias) range where Andreev reflection signatures are observed and the size of the inner gap extracted from the tunneling spectra in \prettyref{fig: fig4}a follow the same trend in the $\nu = -2.15$ to $\nu=-2.4$ range.
So even though the precise interpretation of the Andreev signal depends on details as discussed before\cite{lakePairingSymmetryTwisted2022b,lewandowskiAndreevReflectionSpectroscopy2023a,sainz-cruzJunctionsSuperconductingSymmetry2023,sukhachovAndreevReflectionScanning2023b,biswasAndreevTunnelingSpectroscopy2025},
the close correspondence between the evolution trends of the inner gap and the energy range where an Andreev reflection signal is observed provides strong additional evidence that the two phenomena are linked.

\begin{figure}[p]
\begin{center}
   \includegraphics[width=16cm]{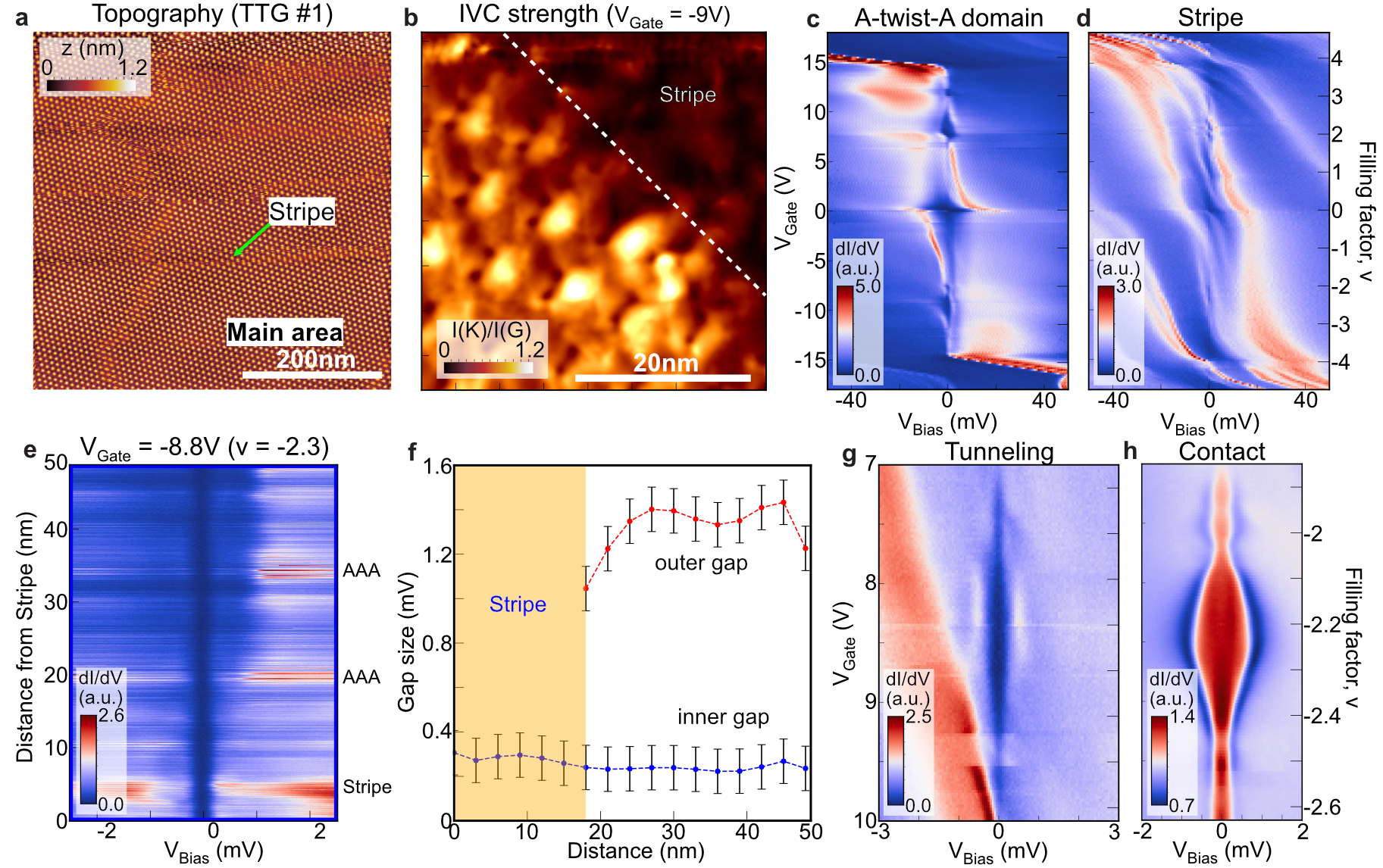}
\end{center}
\caption{{\bf Distinguishing the outer and inner gap on the domain boundary of MATTG.} 
{\bf a}, $500$ nm by $500$ nm size large area topography measured on device $\#1$. Bottom right corner of the topography is where \prettyref{fig: fig3} is measured.
{\bf b}, IVC strength map overlaid on top of $dI/dV$ map displaying a Kekulé pattern
{\bf c, d}, $V_{\rm Gate}$ dependent $dI/dV$ spectroscopy measured on ({\bf d}), and away ({\bf c}) from the stripe domain boundary.
{\bf e}, Tunneling conductance ($dI/dV$) as a function of distance away from the stripe.
{\bf f}, Outer (red) and inner (blue) gap size as a function of distance extracted from \prettyref{fig: fig5}e.
{\bf g}, $V_{\rm Gate}$ dependent $dI/dV$ spectroscopy measured on the center of stripe focused on superconducting gap spectrum.
{\bf h}, Point contact conductance ($dI/dV$) spectroscopy measured on the same spot as \prettyref{fig: fig5}g focused on Andreev reflection signal.
}
\label{fig: fig5}
\end{figure}

Finally, we turn our attention to naturally formed structural anomalies in the MATTG sample and their influence on the two low-temperature gaps. Previous experiments have shown that MATTG locally exhibits mirror-symmetric A-twist-A domains separated by stripe-like domain boundaries \cite{kim_evidence_2022,turkelOrderlyDisorderMagicangle2022,craigLocalAtomicStacking2024}. These boundaries are visible as bright stripe regions in our large-scale image in \prettyref{fig: fig5}a.
Gate-dependent $dI/dV$ spectroscopy at and away from the stripe shows that the cascade of band renormalization is no longer visible on the stripe \prettyref{fig: fig5}c, d. Importantly, on stripes, we observe a smooth crossing of the $dI/dV$ peaks across the Fermi energy without pinning. These results suggest a locally reduced Coulomb on-site interaction, likely resulting from the local change in stacking configuration. Additionally, the inter-valley coherent order, as measured by normalized intensity of the Kekul\'e peaks, is also dramatically suppressed on stripes (by a factor of approximately four, see \prettyref{fig: fig5}b and \prettyref{exfig: Stripe}).

The $dI/dV$ spectrum is also significantly affected by the domain boundaries (\prettyref{fig: fig5}e-g). 
Focusing on $V_{\rm Gate} = -8.8$ V where both inner and outer gaps can be clearly resolved in the A-twist-A bulk we can track the evolution of the gaps as we approach the 
domain boundary (\prettyref{fig: fig5}e). The inner ($\sim 300\mu$eV) gap shows to be 
nearly constant across the boundary (see also \prettyref{fig: fig5}f).  In contrast, the outer 
gap shows more complex evolution: At $50$ nm away from the stripe, we can still identify it
as a broad suppression of the density of states while it completely disappears at about $20$ nm away from the center of the stripe. The high resolution $dI/dV$ spectrum taken in the stripe center confirms real space mapping results. It shows a $300\mu$eV gap pinned to the Fermi level in between filling factors $\nu=-2$ to $\nu=-2.5$ (\prettyref{fig: fig5}g) while the outer gap is 
completely absent. Note that the superconducting nature of the inner gap is further supported by the $dI/dV$ spectrum at the point contact regime that shows the Andreev reflection signal. 
The suppression of the outer gap at the domain boundary, along with the persistence of the inner gap without a reduction in size, highlights the distinct responses of these two gaps to local changes in the band structure. 
The disappearance of the outer gap on the stripe provides another example of how the outer gap follows the emergence of an inter-valley coherent order.

Our study provides the first high-resolution experimental insights on how 
superconductivity and intervalley coherent order in moir\'e graphene systems 
respond to local relaxation and stacking configuration changes. 
Moreover, identifying the nature of the inner and outer gap clarifies several unexplained features observed in previous STM studies\cite{kim_evidence_2022,ohEvidenceUnconventionalSuperconductivity2021}.
The $2\Delta/k_BT_c$ ratio derived from the size of the inner gap is approximately $8.3$ around the filling factor $\nu = -2.3$, which remains significantly larger than the BCS value but is notably smaller than the ratio associated with the pseudogap\cite{yuUniversalRelationshipMagnetic2009a,inosovCrossoverWeakStrong2011b}(\prettyref{exfig: InnerT}). 
More importantly, our measurements offer essential insights into a hierarchy 
of emerging phases in MATTG preceding the superconductivity. 
In particular, our finding that superconductivity and intervalley coherent order are established from 
the Kondo-like parent state has potentially far-reaching consequences\cite{kasuyaElectricalResistanceFerromagnetic1956,doniachKondoLatticeWeak1977,colemanHeavyFermionsElectrons2007}. 
Dynamic correlation effects imply the formation of dispersive heavy bands pinned to Fermi energy.
This offers a framework to interpret recent transport and kinetic inductance studies\cite{tianEvidenceDiracFlat2023,tanakaSuperfluidStiffnessMagicangle2025,banerjeeSuperfluidStiffnessTwisted2025}, based on a mechanism involving the heavy band physics.
Within this picture, the rough estimate of the Fermi velocity of the heavy band is effectively set by the ratio between the IVC gap and moir\'e lengthscale $v_F = \Delta_{IVC}/(\hbar a_{moire}) \sim 10^{4}$ m/s.
While this is consistent with previous experimental values\cite{tianEvidenceDiracFlat2023,tanakaSuperfluidStiffnessMagicangle2025}, it suggests a fundamentally different microscopic origin. 
Moreover, our observation of a transition from a single gap to a double gap structure near $\nu\approx -2.3$ may help connect with 
the transport and kinetic inductance experiments that report distinct superconducting regimes in MATTG\cite{zhouDoubledomeUnconventionalSuperconductivity2024a,mukherjeeSuperconductingMagicangleTwisted2024a,banerjeeSuperfluidStiffnessTwisted2025}.
Importantly, our data highlight the change in the superconducting parent state from a fully gapped (single gap regime) to a shallow V-shaped gapped (double gap regime) spectrum.
The transition in parent state points toward a particular intertwinement of superconductivity and intervalley coherence breaking the Kondo-like phase.
In addition, fully gapped parent state in a single gap regime points to either an unconventional nature of superconductivity\cite{christosNodalBandoffdiagonalSuperconductivity2023a} or possible Bose-Einstein condensation\cite{kim_evidence_2022}.
We envision that future work will address how superconductivity develops within the heavy fermion background set by dynamic correlation and inter-valley coherence\cite{wangMolecularPairingTwisted2024,younHundnessTwistedBilayer2024a}.

\noindent {\bf References:}

\bibliographystyle{naturemag}
\bibliography{ttg_twogap}

\begin{thebibliography}{1}
\expandafter\ifx\csname url\endcsname\relax
  \def\url#1{\texttt{#1}}\fi
\expandafter\ifx\csname urlprefix\endcsname\relax\def\urlprefix{URL }\fi
\providecommand{\bibinfo}[2]{#2}
\providecommand{\eprint}[2][]{\url{#2}}

\bibitem{song2022thf}
\bibinfo{author}{Song, Z.-D.} \& \bibinfo{author}{Bernevig, B.~A.}
\newblock \bibinfo{title}{Magic-angle twisted bilayer graphene as a topological
  heavy fermion problem}.
\newblock \emph{\bibinfo{journal}{Phys. Rev. Lett.}}
  \textbf{\bibinfo{volume}{129}}, \bibinfo{pages}{047601}
  (\bibinfo{year}{2022}).
\newblock
  \urlprefix\url{https://link.aps.org/doi/10.1103/PhysRevLett.129.047601}.

\bibitem{parcolletTRIQST}
\bibinfo{author}{Parcollet, O.} \emph{et~al.}
\newblock \bibinfo{title}{{{TRIQS}}: {{A Toolbox}} for {{Research}} on
  {{Interacting Quantum Systems}}}.
\newblock \emph{\bibinfo{journal}{Computer Physics Communications}}
  \textbf{\bibinfo{volume}{196}}, \bibinfo{pages}{398--415}
  (\bibinfo{year}{2015}).
\newblock \eprint{1504.01952}.

\bibitem{wallerbergerW2dynamics}
\bibinfo{author}{Wallerberger, M.} \emph{et~al.}
\newblock \bibinfo{title}{W2dynamics: {{Local}} one- and two-particle
  quantities from dynamical mean field theory}.
\newblock \emph{\bibinfo{journal}{Computer Physics Communications}}
  \textbf{\bibinfo{volume}{235}}, \bibinfo{pages}{388--399}
  (\bibinfo{year}{2019}).

\bibitem{rai2024-for-si}
\bibinfo{author}{Rai, G.} \emph{et~al.}
\newblock \bibinfo{title}{Dynamical correlations and order in magic-angle
  twisted bilayer graphene}.
\newblock \emph{\bibinfo{journal}{Physical Review X}}
  \textbf{\bibinfo{volume}{14}}, \bibinfo{pages}{031045}
  (\bibinfo{year}{2024}).

\bibitem{hu2023kondo}
\bibinfo{author}{Hu, H.} \emph{et~al.}
\newblock \bibinfo{title}{Symmetric kondo lattice states in doped strained
  twisted bilayer graphene}.
\newblock \emph{\bibinfo{journal}{Phys. Rev. Lett.}}
  \textbf{\bibinfo{volume}{131}}, \bibinfo{pages}{166501}
  (\bibinfo{year}{2023}).
\newblock
  \urlprefix\url{https://link.aps.org/doi/10.1103/PhysRevLett.131.166501}.

\bibitem{horvatic-for-si}
\bibinfo{author}{Horvati{\'c}, B.}, \bibinfo{author}{Sokcevi{\'c}, D.} \&
  \bibinfo{author}{Zlati{\'c}, V.}
\newblock \bibinfo{title}{Finite-temperature spectral density for the
  {{Anderson}} model}.
\newblock \emph{\bibinfo{journal}{Physical Review B}}
  \textbf{\bibinfo{volume}{36}}, \bibinfo{pages}{675--683}
  (\bibinfo{year}{1987}).

\bibitem{kraberger2017maximum}
\bibinfo{author}{Kraberger, G.~J.}, \bibinfo{author}{Triebl, R.},
  \bibinfo{author}{Zingl, M.} \& \bibinfo{author}{Aichhorn, M.}
\newblock \bibinfo{title}{Maximum entropy formalism for the analytic
  continuation of matrix-valued green's functions}.
\newblock \emph{\bibinfo{journal}{Physical Review B}}
  \textbf{\bibinfo{volume}{96}}, \bibinfo{pages}{155128}
  (\bibinfo{year}{2017}).

\end{thebibliography}


\begin{thebibliography}{10}
\expandafter\ifx\csname url\endcsname\relax
  \def\url#1{\texttt{#1}}\fi
\expandafter\ifx\csname urlprefix\endcsname\relax\def\urlprefix{URL }\fi
\providecommand{\bibinfo}[2]{#2}
\providecommand{\eprint}[2][]{\url{#2}}

\bibitem{parkTunableStronglyCoupled2021}
\bibinfo{author}{Park, J.~M.}, \bibinfo{author}{Cao, Y.},
  \bibinfo{author}{Watanabe, K.}, \bibinfo{author}{Taniguchi, T.} \&
  \bibinfo{author}{Jarillo-Herrero, P.}
\newblock \bibinfo{title}{Tunable strongly coupled superconductivity in
  magic-angle twisted trilayer graphene}.
\newblock \emph{\bibinfo{journal}{Nature}} \textbf{\bibinfo{volume}{590}},
  \bibinfo{pages}{249--255} (\bibinfo{year}{2021}).
\newblock \urlprefix\url{https://www.nature.com/articles/s41586-021-03192-0}.
\newblock \bibinfo{note}{Publisher: Nature Publishing Group}.

\bibitem{haoElectricFieldTunable2021}
\bibinfo{author}{Hao, Z.} \emph{et~al.}
\newblock \bibinfo{title}{Electric field{\textendash}tunable superconductivity
  in alternating-twist magic-angle trilayer graphene}.
\newblock \emph{\bibinfo{journal}{Science}} \textbf{\bibinfo{volume}{371}},
  \bibinfo{pages}{1133--1138} (\bibinfo{year}{2021}).
\newblock \urlprefix\url{https://www.science.org/doi/10.1126/science.abg0399}.
\newblock \bibinfo{note}{Publisher: American Association for the Advancement of
  Science}.

\bibitem{parkRobustSuperconductivityMagicangle2022}
\bibinfo{author}{Park, J.~M.} \emph{et~al.}
\newblock \bibinfo{title}{Robust superconductivity in magic-angle multilayer
  graphene family}.
\newblock \emph{\bibinfo{journal}{Nat. Mater.}} \textbf{\bibinfo{volume}{21}},
  \bibinfo{pages}{877--883} (\bibinfo{year}{2022}).
\newblock \urlprefix\url{https://www.nature.com/articles/s41563-022-01287-1}.
\newblock \bibinfo{note}{Publisher: Nature Publishing Group}.

\bibitem{zhang_promotion_2022}
\bibinfo{author}{Zhang, Y.} \emph{et~al.}
\newblock \bibinfo{title}{Promotion of superconductivity in magic-angle
  graphene multilayers}.
\newblock \emph{\bibinfo{journal}{Science}} \textbf{\bibinfo{volume}{377}},
  \bibinfo{pages}{1538--1543} (\bibinfo{year}{2022}).
\newblock \urlprefix\url{https://www.science.org/doi/10.1126/science.abn8585}.

\bibitem{ohEvidenceUnconventionalSuperconductivity2021}
\bibinfo{author}{Oh, M.} \emph{et~al.}
\newblock \bibinfo{title}{Evidence for unconventional superconductivity in
  twisted bilayer graphene}.
\newblock \emph{\bibinfo{journal}{Nature}} \textbf{\bibinfo{volume}{600}},
  \bibinfo{pages}{240--245} (\bibinfo{year}{2021}).
\newblock \urlprefix\url{https://www.nature.com/articles/s41586-021-04121-x}.

\bibitem{kim_evidence_2022}
\bibinfo{author}{Kim, H.} \emph{et~al.}
\newblock \bibinfo{title}{Evidence for unconventional superconductivity in
  twisted trilayer graphene}.
\newblock \emph{\bibinfo{journal}{Nature}} \textbf{\bibinfo{volume}{606}},
  \bibinfo{pages}{494--500} (\bibinfo{year}{2022}).
\newblock \urlprefix\url{https://www.nature.com/articles/s41586-022-04715-z}.
\newblock \bibinfo{note}{Number: 7914 Publisher: Nature Publishing Group}.

\bibitem{songMagicAngleTwistedBilayer2022a}
\bibinfo{author}{Song, Z.-D.} \& \bibinfo{author}{Bernevig, B.~A.}
\newblock \bibinfo{title}{Magic-{Angle} {Twisted} {Bilayer} {Graphene} as a
  {Topological} {Heavy} {Fermion} {Problem}}.
\newblock \emph{\bibinfo{journal}{Phys. Rev. Lett.}}
  \textbf{\bibinfo{volume}{129}}, \bibinfo{pages}{047601}
  (\bibinfo{year}{2022}).
\newblock
  \urlprefix\url{https://link.aps.org/doi/10.1103/PhysRevLett.129.047601}.
\newblock \bibinfo{note}{Publisher: American Physical Society}.

\bibitem{shiHeavyfermionRepresentationTwisted2022}
\bibinfo{author}{Shi, H.} \& \bibinfo{author}{Dai, X.}
\newblock \bibinfo{title}{Heavy-fermion representation for twisted bilayer
  graphene systems}.
\newblock \emph{\bibinfo{journal}{Phys. Rev. B}}
  \textbf{\bibinfo{volume}{106}}, \bibinfo{pages}{245129}
  (\bibinfo{year}{2022}).
\newblock \urlprefix\url{https://link.aps.org/doi/10.1103/PhysRevB.106.245129}.
\newblock \bibinfo{note}{Publisher: American Physical Society}.

\bibitem{suhlDispersionTheoryKondo1965}
\bibinfo{author}{Suhl, H.}
\newblock \bibinfo{title}{Dispersion {Theory} of the {Kondo} {Effect}}.
\newblock \emph{\bibinfo{journal}{Phys. Rev.}} \textbf{\bibinfo{volume}{138}},
  \bibinfo{pages}{A515--A523} (\bibinfo{year}{1965}).
\newblock \urlprefix\url{https://link.aps.org/doi/10.1103/PhysRev.138.A515}.

\bibitem{abrikosovElectronScatteringMagnetic1965}
\bibinfo{author}{Abrikosov, A.~A.}
\newblock \bibinfo{title}{Electron scattering on magnetic impurities in metals
  and anomalous resistivity effects}.
\newblock \emph{\bibinfo{journal}{Physics Physique Fizika}}
  \textbf{\bibinfo{volume}{2}}, \bibinfo{pages}{5--20} (\bibinfo{year}{1965}).
\newblock
  \urlprefix\url{https://link.aps.org/doi/10.1103/PhysicsPhysiqueFizika.2.5}.

\bibitem{saitoIsospinPomeranchukEffect2021}
\bibinfo{author}{Saito, Y.} \emph{et~al.}
\newblock \bibinfo{title}{Isospin {Pomeranchuk} effect in twisted bilayer
  graphene}.
\newblock \emph{\bibinfo{journal}{Nature}} \textbf{\bibinfo{volume}{592}},
  \bibinfo{pages}{220--224} (\bibinfo{year}{2021}).
\newblock \urlprefix\url{https://www.nature.com/articles/s41586-021-03409-2}.
\newblock \bibinfo{note}{Publisher: Nature Publishing Group}.

\bibitem{zondinerCascadePhaseTransitions2020b}
\bibinfo{author}{Zondiner, U.} \emph{et~al.}
\newblock \bibinfo{title}{Cascade of phase transitions and {Dirac} revivals in
  magic-angle graphene}.
\newblock \emph{\bibinfo{journal}{Nature}} \textbf{\bibinfo{volume}{582}},
  \bibinfo{pages}{203--208} (\bibinfo{year}{2020}).
\newblock \urlprefix\url{https://www.nature.com/articles/s41586-020-2373-y}.
\newblock \bibinfo{note}{Publisher: Nature Publishing Group}.

\bibitem{rozenEntropicEvidencePomeranchuk2021}
\bibinfo{author}{Rozen, A.} \emph{et~al.}
\newblock \bibinfo{title}{Entropic evidence for a {Pomeranchuk} effect in
  magic-angle graphene}.
\newblock \emph{\bibinfo{journal}{Nature}} \textbf{\bibinfo{volume}{592}},
  \bibinfo{pages}{214--219} (\bibinfo{year}{2021}).
\newblock \urlprefix\url{https://www.nature.com/articles/s41586-021-03319-3}.
\newblock \bibinfo{note}{Publisher: Nature Publishing Group}.

\bibitem{merinoEvidenceHeavyFermion2024a}
\bibinfo{author}{Merino, R.~L.} \emph{et~al.}
\newblock \bibinfo{title}{Evidence of heavy fermion physics in the
  thermoelectric transport of magic angle twisted bilayer graphene}
  (\bibinfo{year}{2024}).
\newblock \urlprefix\url{http://arxiv.org/abs/2402.11749}.
\newblock \bibinfo{note}{ArXiv:2402.11749 [cond-mat]}.

\bibitem{batlle-porroCryoNearFieldPhotovoltageMicroscopy2024}
\bibinfo{author}{Batlle-Porro, S.} \emph{et~al.}
\newblock \bibinfo{title}{Cryo-{Near}-{Field} {Photovoltage} {Microscopy} of
  {Heavy}-{Fermion} {Twisted} {Symmetric} {Trilayer} {Graphene}}
  (\bibinfo{year}{2024}).
\newblock \urlprefix\url{http://arxiv.org/abs/2402.12296}.
\newblock \bibinfo{note}{ArXiv:2402.12296 [cond-mat]}.

\bibitem{ghoshThermopowerProbesEmergent2025}
\bibinfo{author}{Ghosh, A.} \emph{et~al.}
\newblock \bibinfo{title}{Thermopower probes of emergent local moments in
  magic-angle twisted bilayer graphene}.
\newblock \emph{\bibinfo{journal}{Nat. Phys.}} \bibinfo{pages}{1--8}
  (\bibinfo{year}{2025}).
\newblock \urlprefix\url{https://www.nature.com/articles/s41567-025-02849-1}.
\newblock \bibinfo{note}{Publisher: Nature Publishing Group}.

\bibitem{calugaruThermoelectricEffectIts2024}
\bibinfo{author}{C{\u a}lug{\u a}ru, D.} \emph{et~al.}
\newblock \bibinfo{title}{The {Thermoelectric} {Effect} and {Its} {Natural}
  {Heavy} {Fermion} {Explanation} in {Twisted} {Bilayer} and {Trilayer}
  {Graphene}} (\bibinfo{year}{2024}).
\newblock \urlprefix\url{http://arxiv.org/abs/2402.14057}.
\newblock \bibinfo{note}{ArXiv:2402.14057 [cond-mat]}.

\bibitem{khalafMagicAngleHierarchy2019b}
\bibinfo{author}{Khalaf, E.}, \bibinfo{author}{Kruchkov, A.~J.},
  \bibinfo{author}{Tarnopolsky, G.} \& \bibinfo{author}{Vishwanath, A.}
\newblock \bibinfo{title}{Magic angle hierarchy in twisted graphene
  multilayers}.
\newblock \emph{\bibinfo{journal}{Phys. Rev. B}}
  \textbf{\bibinfo{volume}{100}}, \bibinfo{pages}{085109}
  (\bibinfo{year}{2019}).
\newblock \urlprefix\url{https://link.aps.org/doi/10.1103/PhysRevB.100.085109}.

\bibitem{leiMirrorSymmetryBreaking2021}
\bibinfo{author}{Lei, C.}, \bibinfo{author}{Linhart, L.}, \bibinfo{author}{Qin,
  W.}, \bibinfo{author}{Libisch, F.} \& \bibinfo{author}{MacDonald, A.~H.}
\newblock \bibinfo{title}{Mirror symmetry breaking and lateral stacking shifts
  in twisted trilayer graphene}.
\newblock \emph{\bibinfo{journal}{Phys. Rev. B}}
  \textbf{\bibinfo{volume}{104}}, \bibinfo{pages}{035139}
  (\bibinfo{year}{2021}).
\newblock \urlprefix\url{https://link.aps.org/doi/10.1103/PhysRevB.104.035139}.
\newblock \bibinfo{note}{Publisher: American Physical Society}.

\bibitem{carrUltraheavyUltrarelativisticDirac2020a}
\bibinfo{author}{Carr, S.} \emph{et~al.}
\newblock \bibinfo{title}{Ultraheavy and {Ultrarelativistic} {Dirac}
  {Quasiparticles} in {Sandwiched} {Graphenes}}.
\newblock \emph{\bibinfo{journal}{Nano Lett.}} \textbf{\bibinfo{volume}{20}},
  \bibinfo{pages}{3030--3038} (\bibinfo{year}{2020}).
\newblock \urlprefix\url{https://doi.org/10.1021/acs.nanolett.9b04979}.
\newblock \bibinfo{note}{Publisher: American Chemical Society}.

\bibitem{wongCascadeElectronicTransitions2020a}
\bibinfo{author}{Wong, D.} \emph{et~al.}
\newblock \bibinfo{title}{Cascade of electronic transitions in magic-angle
  twisted bilayer graphene}.
\newblock \emph{\bibinfo{journal}{Nature}} \textbf{\bibinfo{volume}{582}},
  \bibinfo{pages}{198--202} (\bibinfo{year}{2020}).
\newblock \urlprefix\url{https://www.nature.com/articles/s41586-020-2339-0}.
\newblock \bibinfo{note}{Publisher: Nature Publishing Group}.

\bibitem{kerelskyMaximizedElectronInteractions2019b}
\bibinfo{author}{Kerelsky, A.} \emph{et~al.}
\newblock \bibinfo{title}{Maximized electron interactions at the magic angle in
  twisted bilayer graphene}.
\newblock \emph{\bibinfo{journal}{Nature}} \textbf{\bibinfo{volume}{572}},
  \bibinfo{pages}{95--100} (\bibinfo{year}{2019}).
\newblock \urlprefix\url{https://www.nature.com/articles/s41586-019-1431-9}.
\newblock \bibinfo{note}{Publisher: Nature Publishing Group}.

\bibitem{choiElectronicCorrelationsTwisted2019b}
\bibinfo{author}{Choi, Y.} \emph{et~al.}
\newblock \bibinfo{title}{Electronic correlations in twisted bilayer graphene
  near the magic angle}.
\newblock \emph{\bibinfo{journal}{Nat. Phys.}} \textbf{\bibinfo{volume}{15}},
  \bibinfo{pages}{1174--1180} (\bibinfo{year}{2019}).
\newblock \urlprefix\url{https://www.nature.com/articles/s41567-019-0606-5}.
\newblock \bibinfo{note}{Publisher: Nature Publishing Group}.

\bibitem{jiangChargeOrderBroken2019a}
\bibinfo{author}{Jiang, Y.} \emph{et~al.}
\newblock \bibinfo{title}{Charge order and broken rotational symmetry in
  magic-angle twisted bilayer graphene}.
\newblock \emph{\bibinfo{journal}{Nature}} \textbf{\bibinfo{volume}{573}},
  \bibinfo{pages}{91--95} (\bibinfo{year}{2019}).
\newblock \urlprefix\url{https://www.nature.com/articles/s41586-019-1460-4}.
\newblock \bibinfo{note}{Publisher: Nature Publishing Group}.

\bibitem{xieSpectroscopicSignaturesManybody2019b}
\bibinfo{author}{Xie, Y.} \emph{et~al.}
\newblock \bibinfo{title}{Spectroscopic signatures of many-body correlations in
  magic-angle twisted bilayer graphene}.
\newblock \emph{\bibinfo{journal}{Nature}} \textbf{\bibinfo{volume}{572}},
  \bibinfo{pages}{101--105} (\bibinfo{year}{2019}).
\newblock \urlprefix\url{https://www.nature.com/articles/s41586-019-1422-x}.
\newblock \bibinfo{note}{Publisher: Nature Publishing Group}.

\bibitem{jiaoMagneticDefectProbes2018}
\bibinfo{author}{Jiao, L.} \emph{et~al.}
\newblock \bibinfo{title}{Magnetic and defect probes of the {SmB6} surface
  state}.
\newblock \emph{\bibinfo{journal}{Science Advances}}
  \textbf{\bibinfo{volume}{4}}, \bibinfo{pages}{eaau4886}
  (\bibinfo{year}{2018}).
\newblock \urlprefix\url{https://www.science.org/doi/10.1126/sciadv.aau4886}.
\newblock \bibinfo{note}{Publisher: American Association for the Advancement of
  Science}.

\bibitem{yeeImagingKondoInsulating2013}
\bibinfo{author}{Yee, M.~M.} \emph{et~al.}
\newblock \bibinfo{title}{Imaging the {Kondo} {Insulating} {Gap} on {SmB6}}
  (\bibinfo{year}{2013}).
\newblock \urlprefix\url{http://arxiv.org/abs/1308.1085}.
\newblock \bibinfo{note}{ArXiv:1308.1085 [cond-mat]}.

\bibitem{seiroEvolutionKondoLattice2018}
\bibinfo{author}{Seiro, S.} \emph{et~al.}
\newblock \bibinfo{title}{Evolution of the {Kondo} lattice and non-{Fermi}
  liquid excitations in a heavy-fermion metal}.
\newblock \emph{\bibinfo{journal}{Nat Commun}} \textbf{\bibinfo{volume}{9}},
  \bibinfo{pages}{3324} (\bibinfo{year}{2018}).
\newblock \urlprefix\url{https://www.nature.com/articles/s41467-018-05801-5}.
\newblock \bibinfo{note}{Publisher: Nature Publishing Group}.

\bibitem{zhangManyBodyResonanceCorrelated2020}
\bibinfo{author}{Zhang, S.~S.} \emph{et~al.}
\newblock \bibinfo{title}{Many-{Body} {Resonance} in a {Correlated}
  {Topological} {Kagome} {Antiferromagnet}}.
\newblock \emph{\bibinfo{journal}{Phys. Rev. Lett.}}
  \textbf{\bibinfo{volume}{125}}, \bibinfo{pages}{046401}
  (\bibinfo{year}{2020}).
\newblock
  \urlprefix\url{https://link.aps.org/doi/10.1103/PhysRevLett.125.046401}.
\newblock \bibinfo{note}{Publisher: American Physical Society}.

\bibitem{horvaticFinitetemperatureSpectralDensity1987}
\bibinfo{author}{Horvati{\'c}, B.}, \bibinfo{author}{Sokcevi{\'c}, D.} \&
  \bibinfo{author}{Zlati{\'c}, V.}
\newblock \bibinfo{title}{Finite-temperature spectral density for the
  {Anderson} model}.
\newblock \emph{\bibinfo{journal}{Phys. Rev. B}} \textbf{\bibinfo{volume}{36}},
  \bibinfo{pages}{675--683} (\bibinfo{year}{1987}).
\newblock \urlprefix\url{https://link.aps.org/doi/10.1103/PhysRevB.36.675}.

\bibitem{raiDynamicalCorrelationsOrder2024}
\bibinfo{author}{Rai, G.} \emph{et~al.}
\newblock \bibinfo{title}{Dynamical {Correlations} and {Order} in
  {Magic}-{Angle} {Twisted} {Bilayer} {Graphene}}.
\newblock \emph{\bibinfo{journal}{Phys. Rev. X}} \textbf{\bibinfo{volume}{14}},
  \bibinfo{pages}{031045} (\bibinfo{year}{2024}).
\newblock \urlprefix\url{https://link.aps.org/doi/10.1103/PhysRevX.14.031045}.

\bibitem{yuMagicangleTwistedSymmetric2023}
\bibinfo{author}{Yu, J.}, \bibinfo{author}{Xie, M.}, \bibinfo{author}{Bernevig,
  B.~A.} \& \bibinfo{author}{Das~Sarma, S.}
\newblock \bibinfo{title}{Magic-angle twisted symmetric trilayer graphene as a
  topological heavy-fermion problem}.
\newblock \emph{\bibinfo{journal}{Phys. Rev. B}}
  \textbf{\bibinfo{volume}{108}}, \bibinfo{pages}{035129}
  (\bibinfo{year}{2023}).
\newblock \urlprefix\url{https://link.aps.org/doi/10.1103/PhysRevB.108.035129}.
\newblock \bibinfo{note}{Publisher: American Physical Society}.

\bibitem{kwanKekuleSpiralOrder2021a}
\bibinfo{author}{Kwan, Y.~H.} \emph{et~al.}
\newblock \bibinfo{title}{Kekul{\textbackslash}'e {Spiral} {Order} at {All}
  {Nonzero} {Integer} {Fillings} in {Twisted} {Bilayer} {Graphene}}.
\newblock \emph{\bibinfo{journal}{Phys. Rev. X}} \textbf{\bibinfo{volume}{11}},
  \bibinfo{pages}{041063} (\bibinfo{year}{2021}).
\newblock \urlprefix\url{https://link.aps.org/doi/10.1103/PhysRevX.11.041063}.
\newblock \bibinfo{note}{Publisher: American Physical Society}.

\bibitem{herzog-arbeitmanHeavyFermionsEfficient2025}
\bibinfo{author}{Herzog-Arbeitman, J.} \emph{et~al.}
\newblock \bibinfo{title}{Heavy {Fermions} as an {Efficient} {Representation}
  of {Atomistic} {Strain} and {Relaxation} in {Twisted} {Bilayer} {Graphene}}
  (\bibinfo{year}{2025}).
\newblock \urlprefix\url{http://arxiv.org/abs/2405.13880}.
\newblock \bibinfo{note}{ArXiv:2405.13880 [cond-mat]}.

\bibitem{herzog-arbeitmanKekuleSpiralOrder2025}
\bibinfo{author}{Herzog-Arbeitman, J.} \emph{et~al.}
\newblock \bibinfo{title}{Kekul{\'e} {Spiral} {Order} from {Strained}
  {Topological} {Heavy} {Fermions}} (\bibinfo{year}{2025}).
\newblock \urlprefix\url{http://arxiv.org/abs/2502.08700}.
\newblock \bibinfo{note}{ArXiv:2502.08700 [cond-mat]}.

\bibitem{bagchiSpinPolarizedScanningTunneling2024}
\bibinfo{author}{Bagchi, M.} \emph{et~al.}
\newblock \bibinfo{title}{Spin-{Polarized} {Scanning} {Tunneling} {Microscopy}
  {Measurements} of an {Anderson} {Impurity}}.
\newblock \emph{\bibinfo{journal}{Phys. Rev. Lett.}}
  \textbf{\bibinfo{volume}{133}}, \bibinfo{pages}{246701}
  (\bibinfo{year}{2024}).
\newblock
  \urlprefix\url{https://link.aps.org/doi/10.1103/PhysRevLett.133.246701}.

\bibitem{hofstetterGeneralizedNumericalRenormalization2000}
\bibinfo{author}{Hofstetter, W.}
\newblock \bibinfo{title}{Generalized {Numerical} {Renormalization} {Group} for
  {Dynamical} {Quantities}}.
\newblock \emph{\bibinfo{journal}{Phys. Rev. Lett.}}
  \textbf{\bibinfo{volume}{85}}, \bibinfo{pages}{1508--1511}
  (\bibinfo{year}{2000}).
\newblock \urlprefix\url{https://link.aps.org/doi/10.1103/PhysRevLett.85.1508}.

\bibitem{kim_imaging_2023}
\bibinfo{author}{Kim, H.} \emph{et~al.}
\newblock \bibinfo{title}{Imaging inter-valley coherent order in magic-angle
  twisted trilayer graphene}.
\newblock \emph{\bibinfo{journal}{Nature}} \textbf{\bibinfo{volume}{623}},
  \bibinfo{pages}{942--948} (\bibinfo{year}{2023}).
\newblock \urlprefix\url{https://www.nature.com/articles/s41586-023-06663-8}.
\newblock \bibinfo{note}{Number: 7989 Publisher: Nature Publishing Group}.

\bibitem{deutscherAndreevSaintJamesReflectionsProbe2005}
\bibinfo{author}{Deutscher, G.}
\newblock \bibinfo{title}{Andreev--{Saint}-{James} reflections: {A} probe of
  cuprate superconductors}.
\newblock \emph{\bibinfo{journal}{Rev. Mod. Phys.}}
  \textbf{\bibinfo{volume}{77}}, \bibinfo{pages}{109--135}
  (\bibinfo{year}{2005}).
\newblock \urlprefix\url{https://link.aps.org/doi/10.1103/RevModPhys.77.109}.
\newblock \bibinfo{note}{Publisher: American Physical Society}.

\bibitem{blonderTransitionMetallicTunneling1982b}
\bibinfo{author}{Blonder, G.~E.}, \bibinfo{author}{Tinkham, M.} \&
  \bibinfo{author}{Klapwijk, T.~M.}
\newblock \bibinfo{title}{Transition from metallic to tunneling regimes in
  superconducting microconstrictions: {Excess} current, charge imbalance, and
  supercurrent conversion}.
\newblock \emph{\bibinfo{journal}{Phys. Rev. B}} \textbf{\bibinfo{volume}{25}},
  \bibinfo{pages}{4515--4532} (\bibinfo{year}{1982}).
\newblock \urlprefix\url{https://link.aps.org/doi/10.1103/PhysRevB.25.4515}.

\bibitem{lakePairingSymmetryTwisted2022b}
\bibinfo{author}{Lake, E.}, \bibinfo{author}{Patri, A.~S.} \&
  \bibinfo{author}{Senthil, T.}
\newblock \bibinfo{title}{Pairing symmetry of twisted bilayer graphene: {A}
  phenomenological synthesis}.
\newblock \emph{\bibinfo{journal}{Phys. Rev. B}}
  \textbf{\bibinfo{volume}{106}}, \bibinfo{pages}{104506}
  (\bibinfo{year}{2022}).
\newblock \urlprefix\url{https://link.aps.org/doi/10.1103/PhysRevB.106.104506}.

\bibitem{lewandowskiAndreevReflectionSpectroscopy2023a}
\bibinfo{author}{Lewandowski, C.}, \bibinfo{author}{Lantagne-Hurtubise,
  {\'E}.}, \bibinfo{author}{Thomson, A.}, \bibinfo{author}{Nadj-Perge, S.} \&
  \bibinfo{author}{Alicea, J.}
\newblock \bibinfo{title}{Andreev reflection spectroscopy in strongly paired
  superconductors}.
\newblock \emph{\bibinfo{journal}{Phys. Rev. B}}
  \textbf{\bibinfo{volume}{107}}, \bibinfo{pages}{L020502}
  (\bibinfo{year}{2023}).
\newblock
  \urlprefix\url{https://link.aps.org/doi/10.1103/PhysRevB.107.L020502}.

\bibitem{sainz-cruzJunctionsSuperconductingSymmetry2023}
\bibinfo{author}{Sainz-Cruz, H.}, \bibinfo{author}{Pantale{\'o}n, P.~A.},
  \bibinfo{author}{Phong, V.~T.}, \bibinfo{author}{Jimeno-Pozo, A.} \&
  \bibinfo{author}{Guinea, F.}
\newblock \bibinfo{title}{Junctions and {Superconducting} {Symmetry} in
  {Twisted} {Bilayer} {Graphene}}.
\newblock \emph{\bibinfo{journal}{Phys. Rev. Lett.}}
  \textbf{\bibinfo{volume}{131}}, \bibinfo{pages}{016003}
  (\bibinfo{year}{2023}).
\newblock
  \urlprefix\url{https://link.aps.org/doi/10.1103/PhysRevLett.131.016003}.

\bibitem{sukhachovAndreevReflectionScanning2023b}
\bibinfo{author}{Sukhachov, P.~O.}, \bibinfo{author}{Von~Oppen, F.} \&
  \bibinfo{author}{Glazman, L.~I.}
\newblock \bibinfo{title}{Andreev {Reflection} in {Scanning} {Tunneling}
  {Spectroscopy} of {Unconventional} {Superconductors}}.
\newblock \emph{\bibinfo{journal}{Phys. Rev. Lett.}}
  \textbf{\bibinfo{volume}{130}}, \bibinfo{pages}{216002}
  (\bibinfo{year}{2023}).
\newblock
  \urlprefix\url{https://link.aps.org/doi/10.1103/PhysRevLett.130.216002}.

\bibitem{biswasAndreevTunnelingSpectroscopy2025}
\bibinfo{author}{Biswas, S.}, \bibinfo{author}{Suman, S.},
  \bibinfo{author}{Randeria, M.} \& \bibinfo{author}{Sensarma, R.}
\newblock \bibinfo{title}{Andreev versus {Tunneling} {Spectroscopy} of
  {Unconventional} {Flat} {Band} {Superconductors}} (\bibinfo{year}{2025}).
\newblock \urlprefix\url{http://arxiv.org/abs/2503.07744}.
\newblock \bibinfo{note}{ArXiv:2503.07744 [cond-mat]}.

\bibitem{turkelOrderlyDisorderMagicangle2022}
\bibinfo{author}{Turkel, S.} \emph{et~al.}
\newblock \bibinfo{title}{Orderly disorder in magic-angle twisted trilayer
  graphene}.
\newblock \emph{\bibinfo{journal}{Science}} \textbf{\bibinfo{volume}{376}},
  \bibinfo{pages}{193--199} (\bibinfo{year}{2022}).
\newblock \urlprefix\url{https://www.science.org/doi/10.1126/science.abk1895}.
\newblock \bibinfo{note}{Publisher: American Association for the Advancement of
  Science}.

\bibitem{craigLocalAtomicStacking2024}
\bibinfo{author}{Craig, I.~M.} \emph{et~al.}
\newblock \bibinfo{title}{Local atomic stacking and symmetry in twisted
  graphene trilayers}.
\newblock \emph{\bibinfo{journal}{Nat. Mater.}} \textbf{\bibinfo{volume}{23}},
  \bibinfo{pages}{323--330} (\bibinfo{year}{2024}).
\newblock \urlprefix\url{https://www.nature.com/articles/s41563-023-01783-y}.
\newblock \bibinfo{note}{Publisher: Nature Publishing Group}.

\bibitem{yuUniversalRelationshipMagnetic2009a}
\bibinfo{author}{Yu, G.}, \bibinfo{author}{Li, Y.}, \bibinfo{author}{Motoyama,
  E.~M.} \& \bibinfo{author}{Greven, M.}
\newblock \bibinfo{title}{A universal relationship between magnetic resonance
  and superconducting gap in unconventional superconductors}.
\newblock \emph{\bibinfo{journal}{Nature Phys}} \textbf{\bibinfo{volume}{5}},
  \bibinfo{pages}{873--875} (\bibinfo{year}{2009}).
\newblock \urlprefix\url{https://www.nature.com/articles/nphys1426}.
\newblock \bibinfo{note}{Publisher: Nature Publishing Group}.

\bibitem{inosovCrossoverWeakStrong2011b}
\bibinfo{author}{Inosov, D.~S.} \emph{et~al.}
\newblock \bibinfo{title}{Crossover from weak to strong pairing in
  unconventional superconductors}.
\newblock \emph{\bibinfo{journal}{Phys. Rev. B}} \textbf{\bibinfo{volume}{83}},
  \bibinfo{pages}{214520} (\bibinfo{year}{2011}).
\newblock \urlprefix\url{https://link.aps.org/doi/10.1103/PhysRevB.83.214520}.

\bibitem{kasuyaElectricalResistanceFerromagnetic1956}
\bibinfo{author}{Kasuya, T.}
\newblock \bibinfo{title}{Electrical {Resistance} of {Ferromagnetic} {Metals}}.
\newblock \emph{\bibinfo{journal}{Progress of Theoretical Physics}}
  \textbf{\bibinfo{volume}{16}}, \bibinfo{pages}{58--63}
  (\bibinfo{year}{1956}).
\newblock \urlprefix\url{https://doi.org/10.1143/PTP.16.58}.

\bibitem{doniachKondoLatticeWeak1977}
\bibinfo{author}{Doniach, S.}
\newblock \bibinfo{title}{The {Kondo} lattice and weak antiferromagnetism}.
\newblock \emph{\bibinfo{journal}{Physica B+C}} \textbf{\bibinfo{volume}{91}},
  \bibinfo{pages}{231--234} (\bibinfo{year}{1977}).
\newblock
  \urlprefix\url{https://linkinghub.elsevier.com/retrieve/pii/0378436377901905}.

\bibitem{colemanHeavyFermionsElectrons2007}
\bibinfo{author}{Coleman, P.}
\newblock \bibinfo{title}{Heavy {Fermions}: {Electrons} at the {Edge} of
  {Magnetism}}.
\newblock In \emph{\bibinfo{booktitle}{Handbook of {Magnetism} and {Advanced}
  {Magnetic} {Materials}}} (\bibinfo{publisher}{John Wiley \& Sons, Ltd},
  \bibinfo{year}{2007}).
\newblock
  \urlprefix\url{https://onlinelibrary.wiley.com/doi/abs/10.1002/9780470022184.hmm105}.
\newblock \bibinfo{note}{\_eprint:
  https://onlinelibrary.wiley.com/doi/pdf/10.1002/9780470022184.hmm105}.

\bibitem{tianEvidenceDiracFlat2023}
\bibinfo{author}{Tian, H.} \emph{et~al.}
\newblock \bibinfo{title}{Evidence for {Dirac} flat band superconductivity
  enabled by quantum geometry}.
\newblock \emph{\bibinfo{journal}{Nature}} \textbf{\bibinfo{volume}{614}},
  \bibinfo{pages}{440--444} (\bibinfo{year}{2023}).
\newblock \urlprefix\url{https://www.nature.com/articles/s41586-022-05576-2}.
\newblock \bibinfo{note}{Publisher: Nature Publishing Group}.

\bibitem{tanakaSuperfluidStiffnessMagicangle2025}
\bibinfo{author}{Tanaka, M.} \emph{et~al.}
\newblock \bibinfo{title}{Superfluid stiffness of magic-angle twisted bilayer
  graphene}.
\newblock \emph{\bibinfo{journal}{Nature}} \textbf{\bibinfo{volume}{638}},
  \bibinfo{pages}{99--105} (\bibinfo{year}{2025}).
\newblock \urlprefix\url{https://www.nature.com/articles/s41586-024-08494-7}.
\newblock \bibinfo{note}{Publisher: Nature Publishing Group}.

\bibitem{banerjeeSuperfluidStiffnessTwisted2025}
\bibinfo{author}{Banerjee, A.} \emph{et~al.}
\newblock \bibinfo{title}{Superfluid stiffness of twisted trilayer graphene
  superconductors}.
\newblock \emph{\bibinfo{journal}{Nature}} \textbf{\bibinfo{volume}{638}},
  \bibinfo{pages}{93--98} (\bibinfo{year}{2025}).
\newblock \urlprefix\url{https://www.nature.com/articles/s41586-024-08444-3}.
\newblock \bibinfo{note}{Publisher: Nature Publishing Group}.

\bibitem{zhouDoubledomeUnconventionalSuperconductivity2024a}
\bibinfo{author}{Zhou, Z.} \emph{et~al.}
\newblock \bibinfo{title}{Double-dome {Unconventional} {Superconductivity} in
  {Twisted} {Trilayer} {Graphene}} (\bibinfo{year}{2024}).
\newblock \urlprefix\url{http://arxiv.org/abs/2404.09909}.
\newblock \bibinfo{note}{ArXiv:2404.09909 [cond-mat]}.

\bibitem{mukherjeeSuperconductingMagicangleTwisted2024a}
\bibinfo{author}{Mukherjee, A.} \emph{et~al.}
\newblock \bibinfo{title}{Superconducting magic-angle twisted trilayer graphene
  hosts competing magnetic order and moir{\'e} inhomogeneities}
  (\bibinfo{year}{2024}).
\newblock \urlprefix\url{http://arxiv.org/abs/2406.02521}.
\newblock \bibinfo{note}{ArXiv:2406.02521 [cond-mat]}.

\bibitem{christosNodalBandoffdiagonalSuperconductivity2023a}
\bibinfo{author}{Christos, M.}, \bibinfo{author}{Sachdev, S.} \&
  \bibinfo{author}{Scheurer, M.~S.}
\newblock \bibinfo{title}{Nodal band-off-diagonal superconductivity in twisted
  graphene superlattices}.
\newblock \emph{\bibinfo{journal}{Nat Commun}} \textbf{\bibinfo{volume}{14}},
  \bibinfo{pages}{7134} (\bibinfo{year}{2023}).
\newblock \urlprefix\url{https://www.nature.com/articles/s41467-023-42471-4}.
\newblock \bibinfo{note}{Publisher: Nature Publishing Group}.

\bibitem{wangMolecularPairingTwisted2024}
\bibinfo{author}{Wang, Y.-J.}, \bibinfo{author}{Zhou, G.-D.},
  \bibinfo{author}{Peng, S.-Y.}, \bibinfo{author}{Lian, B.} \&
  \bibinfo{author}{Song, Z.-D.}
\newblock \bibinfo{title}{Molecular {Pairing} in {Twisted} {Bilayer} {Graphene}
  {Superconductivity}}.
\newblock \emph{\bibinfo{journal}{Phys. Rev. Lett.}}
  \textbf{\bibinfo{volume}{133}}, \bibinfo{pages}{146001}
  (\bibinfo{year}{2024}).
\newblock \urlprefix\url{http://arxiv.org/abs/2402.00869}.
\newblock \bibinfo{note}{ArXiv:2402.00869 [cond-mat]}.

\bibitem{younHundnessTwistedBilayer2024a}
\bibinfo{author}{Youn, S.}, \bibinfo{author}{Goh, B.}, \bibinfo{author}{Zhou,
  G.-D.}, \bibinfo{author}{Song, Z.-D.} \& \bibinfo{author}{Lee, S.-S.~B.}
\newblock \bibinfo{title}{Hundness in twisted bilayer graphene: correlated gaps
  and pairing} (\bibinfo{year}{2024}).
\newblock \urlprefix\url{http://arxiv.org/abs/2412.03108}.
\newblock \bibinfo{note}{ArXiv:2412.03108 [cond-mat]}.

\bibitem{choi_correlation-driven_2021}
\bibinfo{author}{Choi, Y.} \emph{et~al.}
\newblock \bibinfo{title}{Correlation-driven topological phases in magic-angle
  twisted bilayer graphene}.
\newblock \emph{\bibinfo{journal}{Nature}} \textbf{\bibinfo{volume}{589}},
  \bibinfo{pages}{536--541} (\bibinfo{year}{2021}).
\newblock \urlprefix\url{https://www.nature.com/articles/s41586-020-03159-7}.
\newblock \bibinfo{note}{Number: 7843 Publisher: Nature Publishing Group}.

\bibitem{choi_interaction-driven_2021}
\bibinfo{author}{Choi, Y.} \emph{et~al.}
\newblock \bibinfo{title}{Interaction-driven band flattening and correlated
  phases in twisted bilayer graphene}.
\newblock \emph{\bibinfo{journal}{Nat. Phys.}} \textbf{\bibinfo{volume}{17}},
  \bibinfo{pages}{1375--1381} (\bibinfo{year}{2021}).
\newblock \urlprefix\url{https://www.nature.com/articles/s41567-021-01359-0}.

\end{thebibliography}

\clearpage

\clearpage

\vspace{5pt}
\noindent {\bf Methods:}

\vspace{5pt}

\noindent {\bf Device fabrication:} 
The fabrication of the graphite/hBN/TTG device follows a sequential layer pickup process using a poly(bisphenol A carbonate) (PC) film mounted on a polydimethylsiloxane (PDMS) block.
This is done via the standard dry transfer method at temperatures ranging from 60-100 $\degree C$.
The stack then needs to be `flipped' onto a substrate to expose the TTG and electrically contact it without introducing polymer residues. 

MATBG device in \prettyref{fig: fig1} is made with a previously developed PDMS-assisted flipping technique where the sample is directly flipped on top of the PDMS surface\cite{choiElectronicCorrelationsTwisted2019b}.
MATTG device $\#1$ and $\#2$ are made with gold-coated PDMS flipping introduced in our previous paper\cite{kim_imaging_2023}.
Initially, the PC film supporting the stack is detached from the PDMS slide and transferred, with the stack facing downward, onto a PDMS block coated with a Ti/Au ($3/12$ nm) layer deposited through e-beam evaporation. 
The PC layer is then dissolved using N-methyl-2-pyrrolidone (NMP).

To enable electrical connections while maintaining surface cleanliness, a ‘gold stamping’ approach is employed. 
This involves patterning a silicon oxide substrate with a SU8 (SU-8 2005, Microchem) photoresist mold, onto which PDMS (SYLGARD 184, 10:1) is cast and cured. 
Once peeled from the mold, the patterned PDMS stamp undergoes gold deposition (e-beam evaporation, $10$–$20$ nm) before being pressed onto the designated region of the sample at $130\degree C$. 
As only the raised features of the stamp contact the sample, gold is selectively transferred, leaving the remainder of the surface uncontaminated.

\noindent {\bf STM measurements:}
The STM measurements were performed in a Unisoku USM 1300J STM / AFM system using a Platinum/Iridium (Pt/Ir) tip as in our previous works on bilayers \cite{choiElectronicCorrelationsTwisted2019b, choi_correlation-driven_2021,choi_interaction-driven_2021}. 
All reported features are observed with many (usually at least ten) different microtips. 
Unless specified otherwise, data were taken at temperature $T = 400$mK and the parameters for $dI/dV$ spectroscopy measurements were $V_{\rm Bias} = 100$mV and $I = 1$nA, and the lock-in parameters were modulation voltage $V_{\rm mod} = 0.1-1$mV and frequency $f = 973.333$Hz. 
Real space $dI/dV$ maps are taken with the constant height mode (feedback turned off, tilt corrected).
The piezo scanner is calibrated on an Ag(111) crystal and verified by measuring the distance between carbon atoms. 
The twist-angle uncertainty is approximately $\pm0.01\degree$, and is determined by measuring moir\'e wavelengths from topography. 
Filling factor assignment has been performed by identifying features corresponding to full-filling and CNP LDOS suppression\cite{choiElectronicCorrelationsTwisted2019b}.

\noindent {\bf Acknowledgments:}  We thank Jason Alicea, Cyprian Lewandowski, \'Etienne Lantagne-Hurtubise, Alex Thomson, Mohit Randeria, Sayak Biswas, Zhi-da Song, Yi-jie Wang, Geng-Dong Zhou for fruitful discussion.
{\bf Funding:} This work has been primarily
supported by the National Science Foundation (grant no. DMR-2005129) and  
Office of Naval Research (grant no. N142112635). S.N-P. also acknowledge the support of the Institute for 
Quantum Information and Matter, an NSF Physics Frontiers 
Center (PHY-2317110) and the Moore foundation (award 12967). 
H.K. acknowledges support from the Kwanjeong fellowship and the Eddleman Quantum Institute Fellowship. 
L.K. acknowledges support from an IQIM-AWS Quantum postdoctoral fellowship. 
Work at UCSB was supported by the U.S. Department of Energy (Award No. DE-SC0020305) and by the Gordon and Betty Moore Foundation under award GBMF9471. 
This work used facilities supported via the UC Santa Barbara NSF Quantum 
Foundry funded via the Q-AMASE-i program under award DMR-1906325. 
B.A.B. was supported by the Gordon and Betty Moore Foundation through Grant No. GBMF8685 towards the Princeton theory program, the Gordon and Betty Moore Foundation’s EPiQS Initiative (Grant No. GBMF11070), the Office of Naval Research (ONR Grant No. N00014-
20-1-2303), the Global Collaborative Network Grant at
Princeton University, the Simons Investigator Grant No.
404513, the BSF Israel US foundation No. 2018226, the
NSF-MERSEC (Grant No. MERSEC DMR 2011750),
the Simons Collaboration on New Frontiers in Superconductivity, and the Schmidt Foundation at Princeton University. 
H.H. and D.C. were supported by the European Research Council (ERC) under
the European Union's Horizon 2020 research and innovation program (Grant Agreement No. 101020833). G.R., L.C., R.V., G.S., and T.W. acknowledge support from the Deutsche Forschungsgemeinschaft (DFG, German Research Foundation) through QUAST FOR 5249 (Project No. 449872909, projects P4 and P5).
G.S. and L.C. were supported by the W\"urzburg-Dresden Cluster of Excellence on Complexity and Topology in Quantum Matter \textit{ct.qmat} - EXC 2147 (Project No. 390858490).
G.R., L.C., and T.W. acknowledge support from the Cluster of Excellence ‘CUI: Advanced Imaging of Matter' – EXC 2056 (Project No. 390715994), and SPP 2244 (WE 5342/5-1 project No. 422707584).
 L.C. gratefully acknowledges the scientific support and HPC resources provided by the Erlangen National High Performance Computing Center (NHR@FAU) of the Friedrich-Alexander-Universität Erlangen-Nürnberg (FAU) under the NHR project b158cb.
 G.R. gratefully acknowledges the computing time granted by the Resource Allocation Board and provided on the supercomputer Lise and Emmy at NHR@ZIB and NHR@Göttingen as part of the NHR infrastructure (project ID hhp00061).

\noindent {\bf Author Contribution:} H.K. fabricated samples with the
help of Y.C., Y.Z., and L.H., and performed STM measurements. H.K. and S.N.-P.
analyzed the data with the help of L.K. and E.B. 
L.C. and G.R. performed the DMFT calculations and analyzed the results with G. S., R. V., and T.W. S.N.-P. supervised the project. 
H.K. and S.N.-P. wrote the manuscript with input from the other authors.

\noindent {\bf Data availability:} The raw data shown in the main figures are available at 
Zenodo. Other data and code that support the findings of this study are available from the 
corresponding authors upon reasonable request.

\renewcommand{\figurename}{\textbf{Extended Data Fig.}}
\renewcommand{\theHfigure}{Extended.\thefigure}
\setcounter{figure}{0}

\begin{figure}[p]
\begin{center}
   \includegraphics[width=15cm]{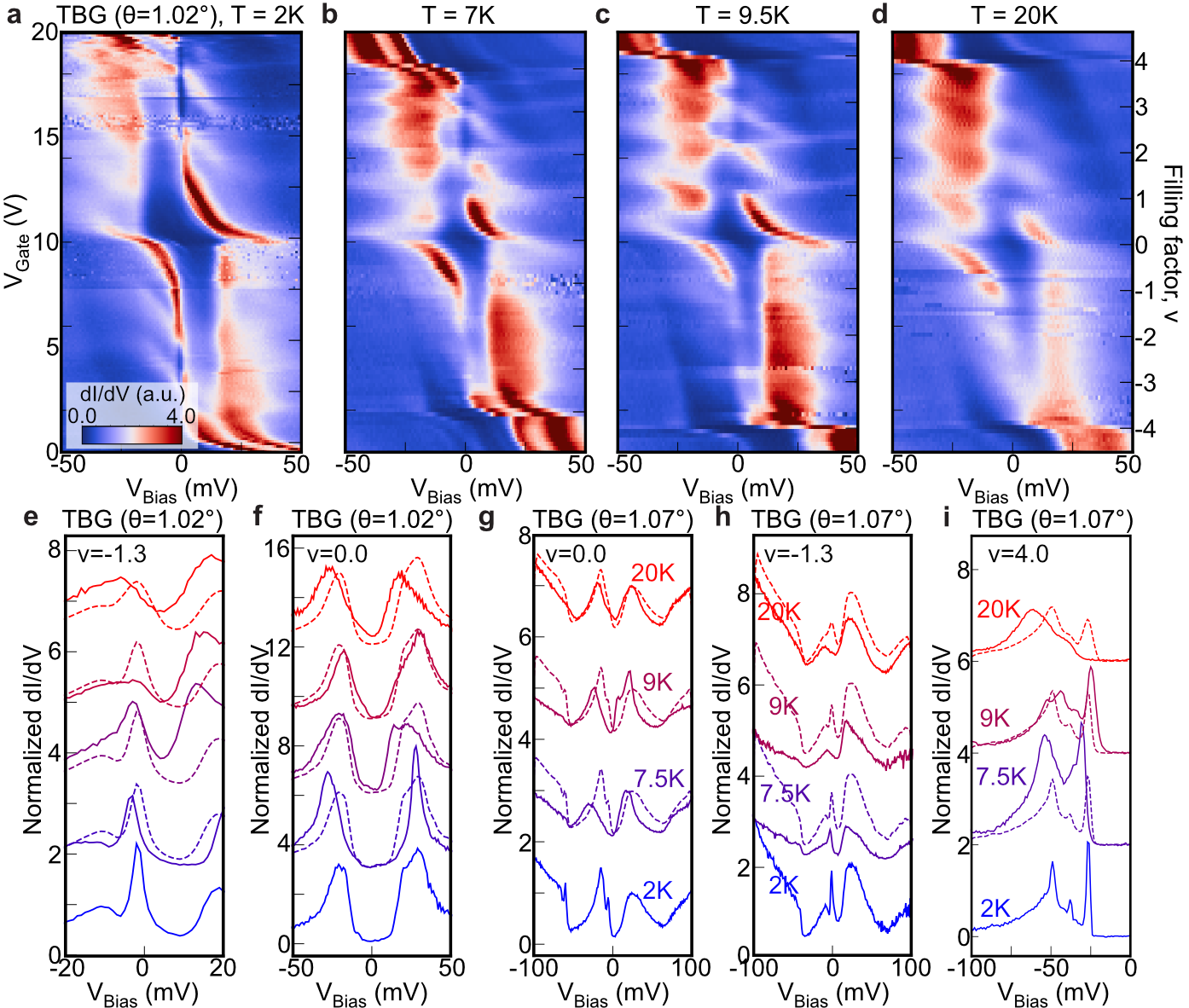}
\end{center}
\caption{{\bf Temperature dependent evolution of MATBG sample.}
{\bf a-d}, $V_{\rm Gate}$ dependent $dI/dV$ spectroscopy measured on MATBG device at an area with local twist angle $\theta = 1.02\degree$ for temperatue $T = 2$ K ({\bf a}), $T = 7$ K ({\bf b}), $T = 9.5$ K ({\bf c}), $T = 20$ K ({\bf d}). 
{\bf e-f}, Temperature dependent spectrum taken from \prettyref{exfig: Temp}a-d at filling factor $\nu = -1.3$ ({\bf e}), $\nu = 0$ ({\bf f}). Dashed line is convolution of $2$ K data with Fermi-Dirac distribution function at each temperature.
{\bf g-i}, Temperature-dependent spectrum taken at different parts of the same MATBG sample where the local twist angle is $\theta = 1.07\degree$ at filling factor $\nu = 0$ ({\bf g}), $\nu = -1.3$ ({\bf h}), $\nu = 5$ ({\bf i}).
}
\label{exfig: Temp}
\end{figure}

\begin{figure}[p]
\begin{center}
   \includegraphics[width=7cm]{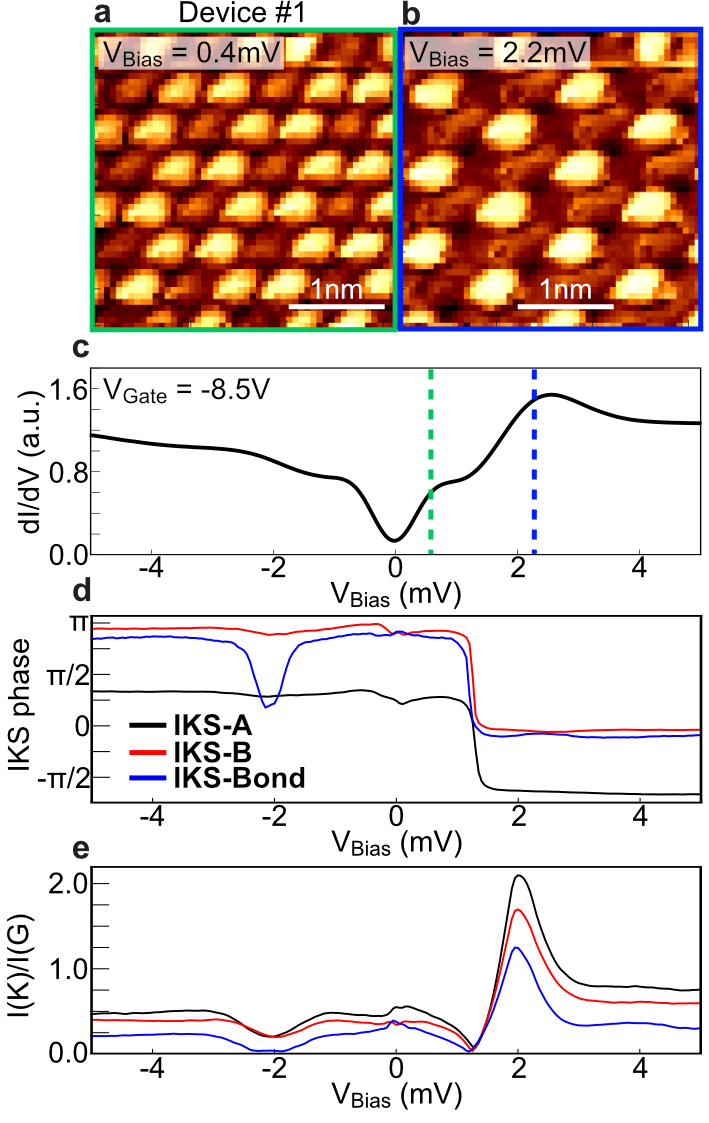}
\end{center}
\caption{{\bf $\mathbf{V_{\rm Bias}}$ dependent mapping of the lattice tripling order on MATTG.}
{\bf a-b}, Real space $dI/dV$ map at $V_{\rm Gate} = -8.5$ V at $V_{\rm Bias} = 0.4$ mV ({\bf a}), $2.2$ mV ({\bf b}). The two real space map shows perfectly inverted Kekul\'e pattern.
{\bf c}, Conductance at $V_{\rm Gate} = -8.5$ V corresponding to a filling factor of $\nu = -2.35$ measured on AAA site on the moire lattice averaged over $3$ nm by $3$ nm region. Green (blue) dashed line corresponds to $V_{\rm Bias}$ where {\bf a} ({\bf b}) is measured.
{\bf d}, Phase of the local IKS order parameter as a function of $V_{\rm Bias}$. Local order parameter decomposition is performed by taking real space $dI/dV$ map for each $V_{\rm Bias}$ and linear transformation of FT Kekul\'e peaks.
{\bf e}, Intensity of the local IKS order parameter as a function of $V_{\rm Bias}$.
}
\label{exfig: Kek_bias}
\end{figure}

\begin{figure}[p]
\begin{center}
   \includegraphics[width=10cm]{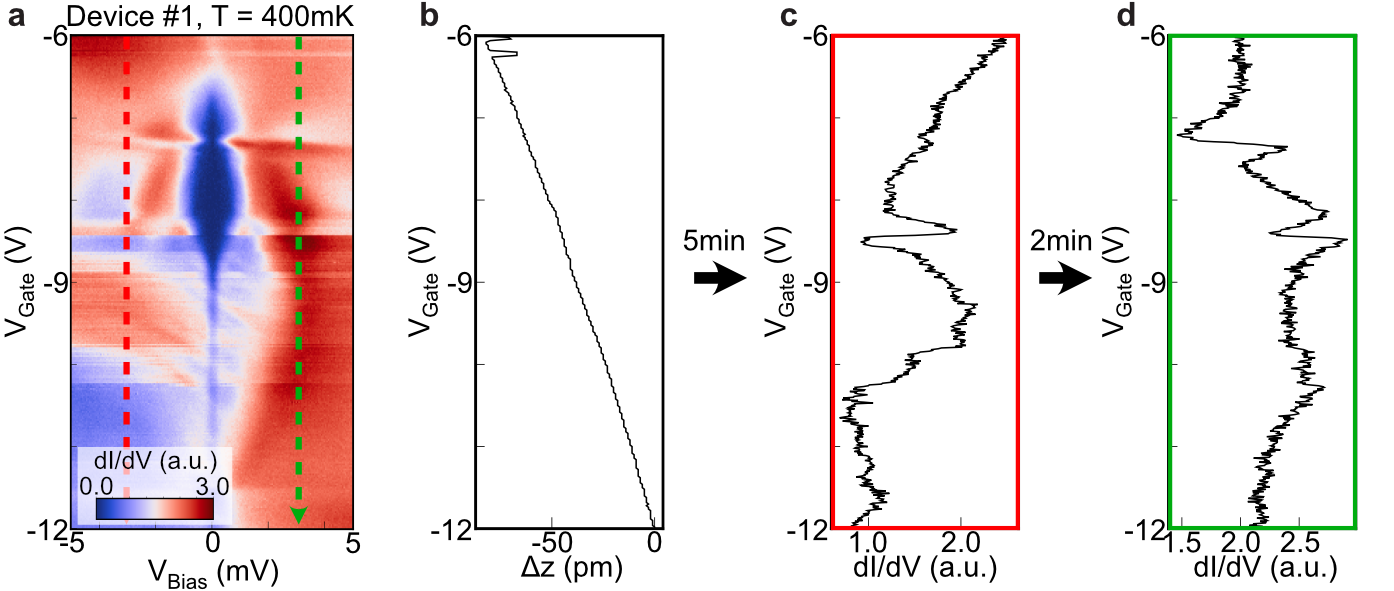}
\end{center}
\caption{{\bf $V_{\rm Gate}$ switching of $dI/dV$ spectrum due to intervalley coherence reconstruction.}
{\bf a}, $V_{\rm Gate}$ dependent $dI/dV$ spectroscopy focusing on hole-doping side where we observe correlated gap. Red (Green) dashed line indicate $V_{\rm Bias}$ which \prettyref{exfig: Switching}c (d) is taken.
{\bf b}, Tip height as a function of $V_{\rm Gate}$ recorded during \prettyref{exfig: Switching}. Tunneling current feedback is turned ON which gives smooth increasing background in height.
{\bf c}, Tunneling $dI/dV$ as a function of $V_{\rm Gate}$ for fixed $V_{\rm Bias} = -3$ mV after $5$ mins taking \prettyref{exfig: Switching}a. Position of $dI/dV$ peak in $V_{\rm Gate}$ matches with the $V_{\rm Gate}$ where switching happens in \prettyref{exfig: Switching}a around $-8.5$ V, indicating that certain swiching behaviors seen in $V_{\rm Gate}$ dependent spectroscopies are reproducible over measurements taken at different time.
{\bf d}, Tunneling $dI/dV$ as a function of $V_{\rm Gate}$ for fixed $V_{\rm Bias} = 3$ mV after $2$ mins taking \prettyref{exfig: Switching}c.
}
\label{exfig: Switching}
\end{figure}

\begin{figure}[p]
\begin{center}
   \includegraphics[width=8cm]{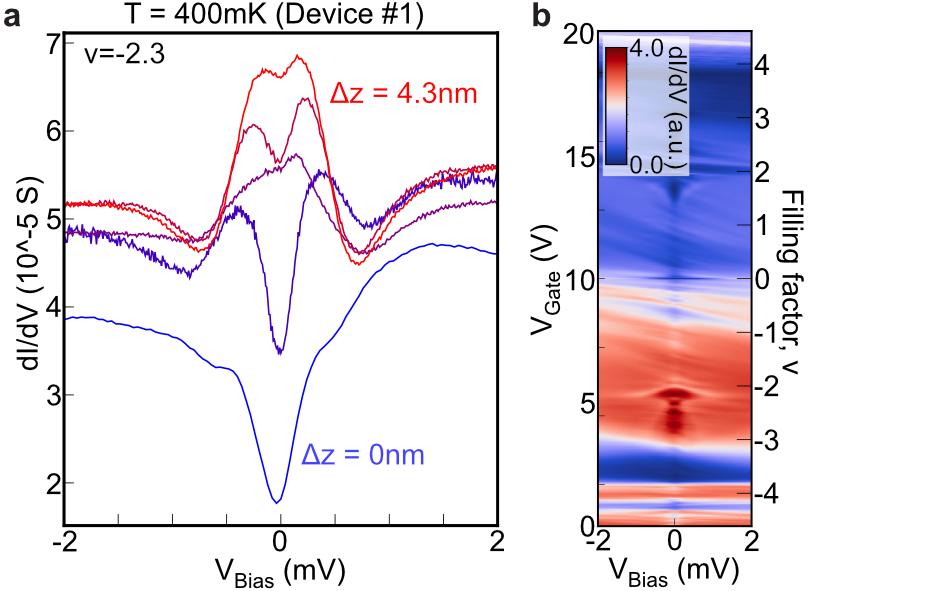}
\end{center}
\caption{{\bf $\Delta z$ dependent $dI/dV$ spectrum tracking the development of Andreev reflection signal.}
{\bf a}, Point-contact $dI/dV$ spectra taken as the tip is approached towards the sample.
{\bf b}, $V_{\rm Gate}$ dependent point contact $dI/dV$ spectroscopy taken on MATTG device $\#1$ at twist angle $\theta = 1.61\degree$. Andreev reflection signal is visible in between filling factor $\nu = -2$ to $-3$.
}
\label{exfig: PCS}
\end{figure}

\begin{figure}[p]
\begin{center}
   \includegraphics[width=14cm]{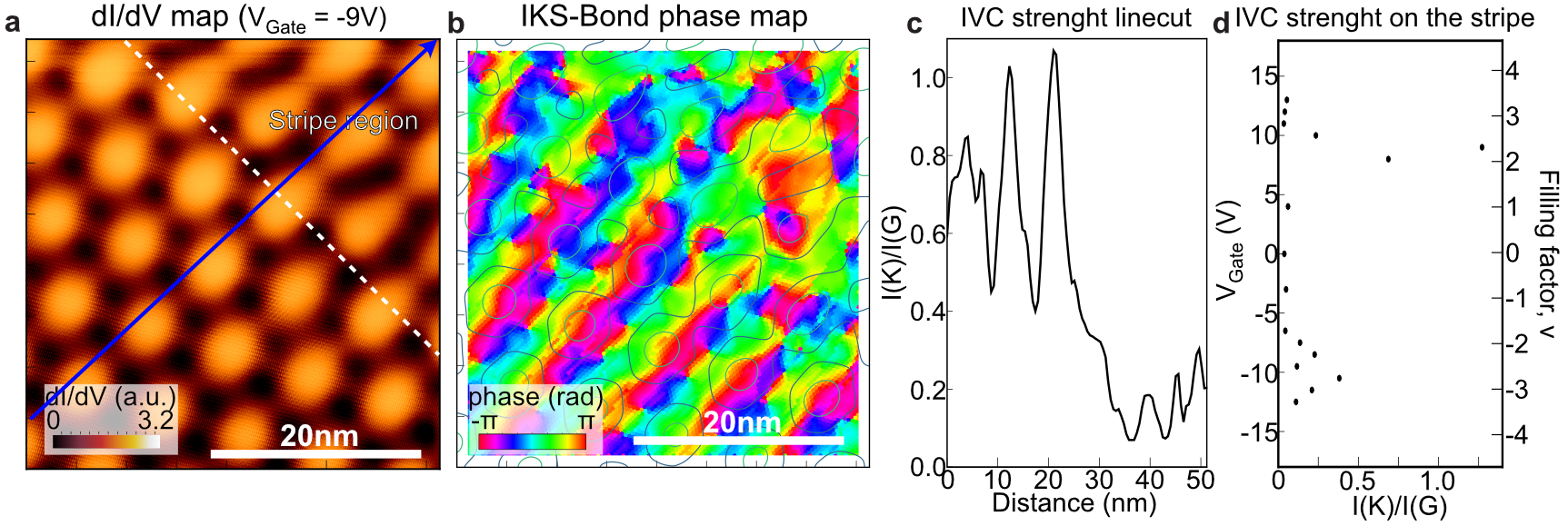}
\end{center}
\caption{{\bf Intensity of FT Kekul\'e peak from real-space $dI/dV$ map adjacent to the MATTG stripe domain boundary.}
{\bf a}, $35$ nm by $35$ nm real-space $dI/dV$ map containing regions on domain boundary and outside of domain boundary. Blue arrow indicate real-space trajectory where the \prettyref{exfig: Stripe}c is extracted.
{\bf b}, Local IKS-Bond order parameter phase extracted from \prettyref{exfig: Stripe}a by taking $2$ nm by $2$ nm window and decomposing Kekul\'e FT peaks into IKS order parameters.
{\bf c}, Intensity of FT Kekul\'e peak normalized by FT graphene peak extracted in real space as one moves towards the stripe domain boundary.
{\bf d}, Intensity of FT Kekul\'e peak normalized by FT graphene peak measured at the center of the domain boundary as a function of $V_{\rm Gate}$.
}
\label{exfig: Stripe}
\end{figure}

\begin{figure}[p]
\begin{center}
   \includegraphics[width=10cm]{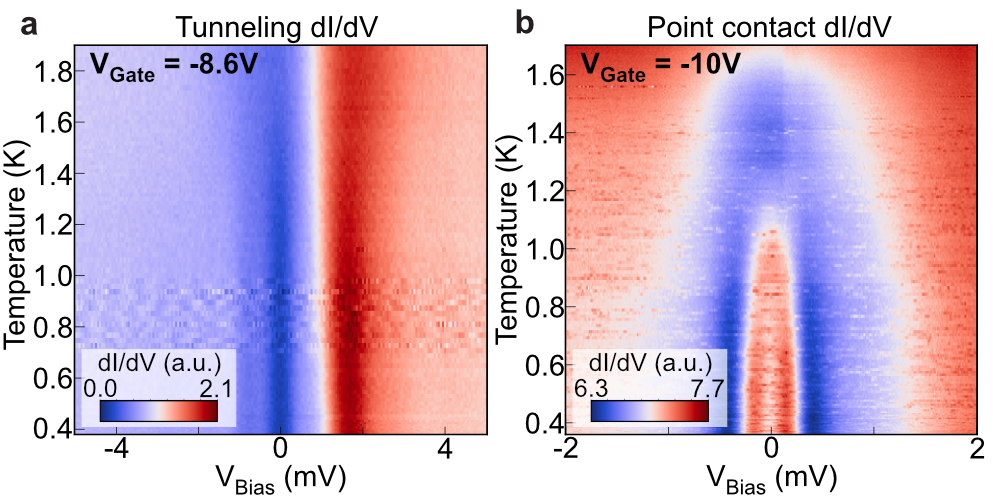}
\end{center}
\caption{{\bf Onset temperature of inner gap signal in tunneling $dI/dV$ and point contact spectroscopy in MATTG device $\#1$.}
{\bf a}, Tunneling $dI/dV$ spectra at fixed $V_{\rm Gate} = -8.6$ V measured as a function of temperature on AAA site.
{\bf b}, Temperature dependence of point contact $dI/dV$ spectra at $V_{\rm Gate} = -10$ V.
}
\label{exfig: InnerT}
\end{figure}

\begin{figure}[p]
\begin{center}
   \includegraphics[width=14cm]{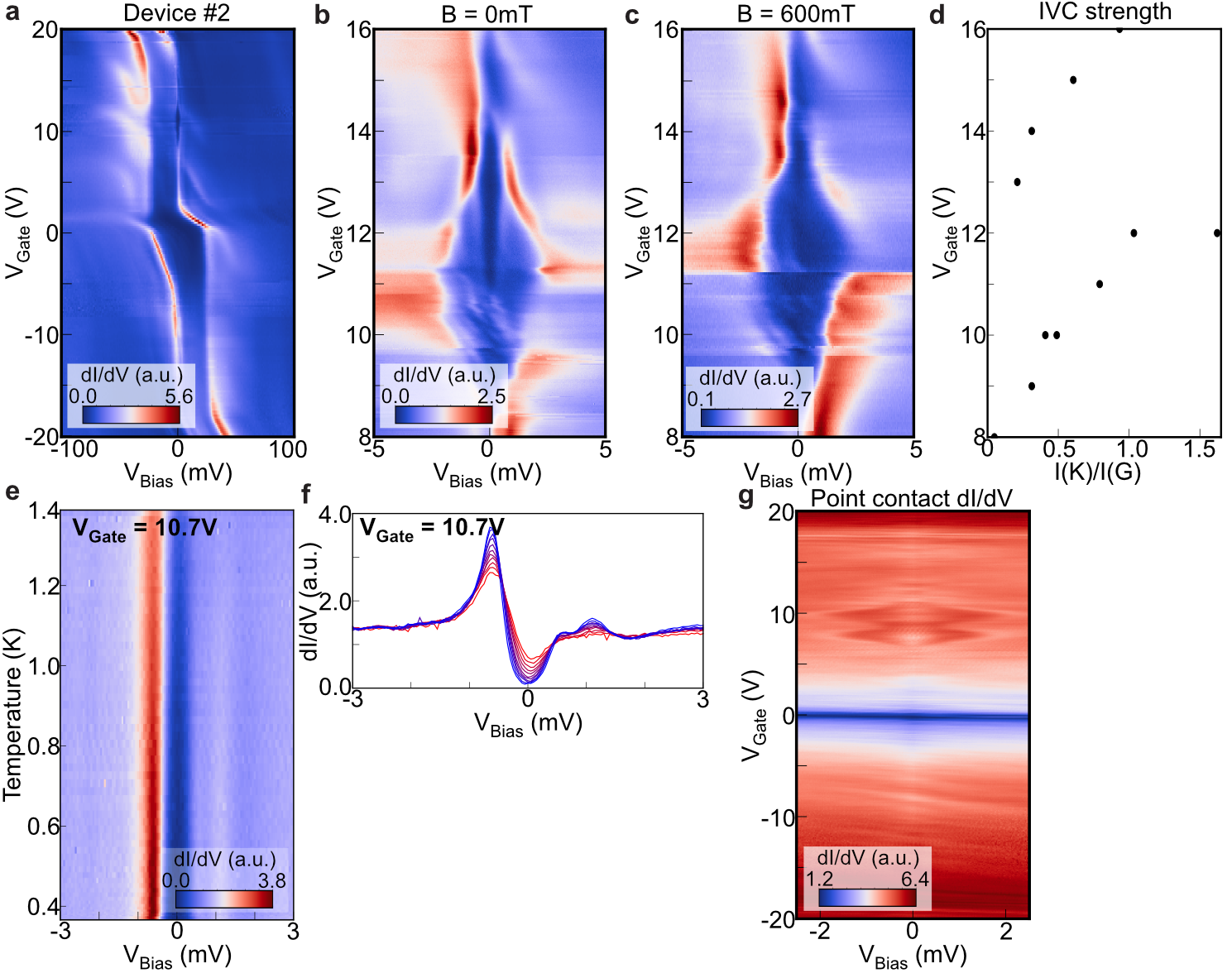}
\end{center}
\caption{{\bf Two gap $dI/dV$ spectrum characterized on MATTG Device $\#2$ $\theta = 1.38\degree$.}
{\bf a}, $V_{\rm Gate}$ dependent $dI/dV$ spectrum taken at $T = 400$ mK on AAA site featuring correlated gaps on electron-doped side.
{\bf b, c}, $V_{\rm Gate}$ dependent $dI/dV$ spectrum focusing in between $\nu = 2$ to $3$ at out-of-plane magnetic field $0$ T ({\bf b}), and $600$ mT ({\bf c}).
{\bf d}, Intensity of the peaks at Kekul\'e reciprocal lattice vector normalized by the intensity of 
the peaks at graphene reciprocal lattice vector as a function of $V_{\rm Gate}$.
$V_{\rm Bias}$ where the real-space $dI/dV$ map is taken is chosen to be at the coherence peak of the gap.
{\bf e}, Tunneling $dI/dV$ spectrum as a function of temperature at fixed $V_{\rm Gate} = 10.7$ V.
{\bf f}, 
{\bf g}, Point-contact $dI/dV$ spectrum showing weak Andreev reflection signal on the electron-side.
}
\label{exfig: TTG16}
\end{figure}

\begin{figure}[p]
\begin{center}
   \includegraphics[width=14cm]{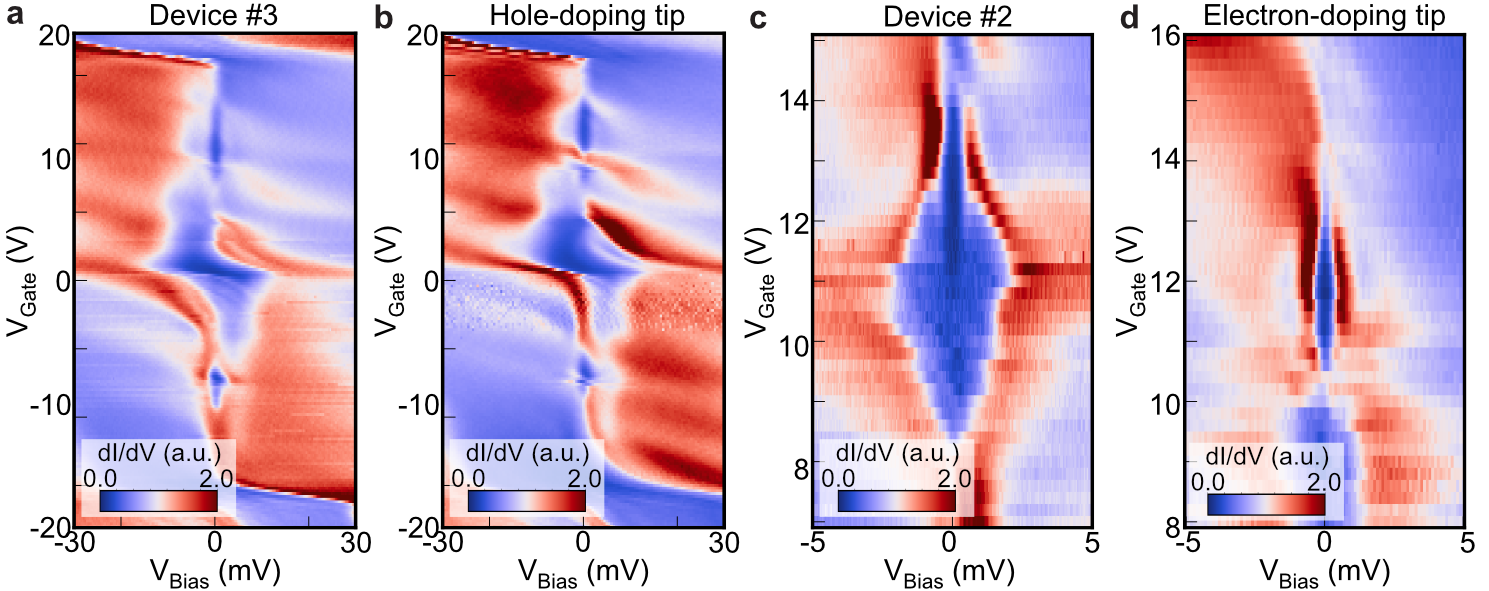}
\end{center}
\caption{{\bf Local doping of the correlated gaps due to work function mis-matching STM tips}
{\bf a-b}, $V_{\rm Gate}$ dependent $dI/dV$ spectrum on area $\theta = 1.51\degree$ with STM tips no local doping ({\bf a}), and local hole doping ({\bf b}). Cascade features are moved to positive $V_{\rm Gate}$ for hole doping STM tips. 
{\bf c-d}, $V_{\rm Gate}$ dependent $dI/dV$ spectrum on area $\theta = 1.38\degree$ with STM tips no local doping ({\bf c}), and local electron doping ({\bf d}). Outer gap is moved to negative $V_{\rm Gate}$ for electron doping STM tips while minimal shift is observed for inner gap.
}
\label{exfig: Tip}
\end{figure}

\clearpage

\beginSI

\noindent \textbf{\large Supplementary Information: Resolving Intervalley Gaps and Many-Body Resonances in Moir\'e Superconductor} 

\section{Dynamical mean-field theory simulations and analysis}
For theoretical simulations on twisted bilayer graphene, we used the Topological Heavy-Fermion (THF) model~\citeSM{song2022thf}. This model consists of 2 localized Wannier orbitals ($f$-orbitals in the following) and 4 dispersive ($c$) modes per valley per spin. It faithfully reproduces the low energy sector of the Bistritzer-MacDonald model and the symmetry representations of the wavefunctions at the high-symmetry points of the moir\'e Brillouin zone.
This model, featuring correlated orbitals hybridizing with dispersive orbitals, belongs to the family of Periodic Anderson Models (PAMs), albeit with a more complicated momentum-dependent hybridization than what is usually considered.
These models are ideal candidates to be solved in the framework of Dynamical Mean-Field Theory (DMFT). We used Continuous-Time Quantum Monte Carlo (CT-QMC) solvers: we employed the TRIQS software suite~\citeSM{parcolletTRIQST} in the symmetry-broken phase, and the w2dynamics solver~\citeSM{wallerbergerW2dynamics} in the symmetric phase. More details on the practical implementation of the calculations can be found in~\citeSM{rai2024-for-si}. The THF model in the absence of strain is perfectly particle-hole symmetric around the charge neutrality point. The data in Fig. 2c, d of the main text were obtained for hole-doping, $\nu<0$, and particle-hole transformed. 

\subsection{Intermediate valence picture for the zero-bias resonance}
\begin{figure}[ht]
    \centering
    \includegraphics[width=\textwidth]{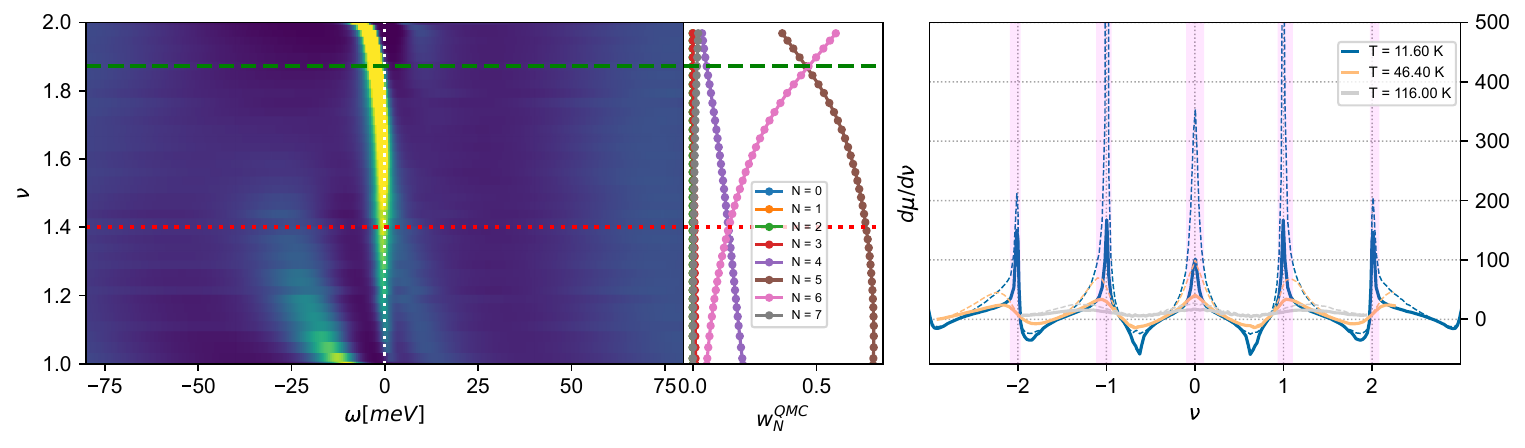}
    \caption{Left: Spectral weight distribution in the doping range $\nu \in [1,2]$ for $T=11.6$ K (left), and the corresponding statistical weight of the occupation sectors extracted from QMC (right). The vertical lines highlight the two relevant occupations at which the relative spectral weight of different sectors switch.
    Right: inverse total (solid lines) and $f$-orbital (dashed lines) charge compressibility for three values of temperature, as a function of the total doping. The shaded region represent the doping ranges where the system can be considered incompressible. Data from~\protect\citeSM{rai2024-for-si}.}
    \label{sifig:compressibility}
\end{figure}

Both experimental observation and theoretical DMFT modeling show the presence of a zero-bias spectral peak, which is rather robust across a wide range of fillings. Here we address the question on whether this is of Kondo origin or representative of quasiparticle-like excitations in an intermediate valence regime. 

\prettyref{sifig:compressibility} (left) shows the spectral function in the symmetric phase, obtained via a CT-QMC simulation at  $T=11.6$ K. This shows a persistent central resonance, which remains pinned at zero frequency in a large doping range but is missing around integer filling. In order to explain the physical origin of this resonance, we turn to the inverse charge compressibility $\frac{\partial\mu}{\partial \nu}$ of the system, plotted on the right panel. This is a measure of how much the system suppresses charge fluctuations: when it diverges, the system is incompressible and charge fluctuations are suppressed. In the plot, the solid lines refer to the \textit{total} charge compressibility, including contributions from the correlated and uncorrelated subspaces, while the dashed lines refer to the compressibility of the correlated subspace alone.
They show the same qualitative behavior, which means the experimentally accessible total charge compressibility is a good indicator of the underlying physics of the narrow $f$-bands. 

The behavior of compressibilities as a function of temperature has been previously analyzed in~\citeSM{rai2024-for-si}, and it shows a remarkable dependence on filling and temperature. Let us first consider the blue lines, referring to a rather low temperature of $T=11.6K$. Here, the system becomes incompressible in a small doping range around \textit{total} integer fillings. In the right panel of \prettyref{sifig:compressibility}, this is indicatively represented via the shaded regions. This scenario, where charge fluctuations are suppressed, would be the most likely to feature a resonance due to the Kondo effect. Yet, due to the strong suppression of local moments~\citeSM{hu2023kondo,rai2024-for-si}, no such resonance arises.

By contrast, away from the $\frac{\partial\mu}{\partial\nu}$ peaks, the system is compressible, which entails rather large charge fluctuations. This is the doping range where the zero-bias resonance manifests. Such behavior is fundamentally at odds with a Kondo-like scenario: for Kondo screening to be the cause of the resonance, local moments would need to develop in the correlated subspace, which would require strong charge fluctuation suppression. Hence, the origin of the resonance is likely to be found somewhere else, and specifically in mixed-valence physics~\citeSM{horvatic-for-si}.

This picture is also supported by the spectral weight distribution itself: this clearly consists of a central peak and two asymmetric Hubbard bands. The central peak emerges from one of the Hubbard bands at fractional filling.

Let us now turn to the orange and grey curves, representing data at medium to high temperatures, $T=46.4K$ and $T=116K$. Here, the compressibility peaks are thermally washed away. Concordantly, there is no clear differentiation between integer and fractional fillings, and the spectral weight distribution does not show a clearly pinned peak anymore (cfr. \prettyref{fig: fig1}h).

To support our determination of mixed-valence physics as the cause of the resonance, we can make use of a convenient byproduct of the DMFT calculations: since in our simulations we sample the partition function $Z$ at temperature $T$, we can access the Boltzmann weight $w^{QMC}_{N}$ of various electronic eigenstates contributing to it, which feature a different occupation of the correlated orbitals from completely empty ($N=0$) to completely full ($N=8$). These weights are plotted, as a function of the occupation, in the side panel of~\prettyref{sifig:compressibility}. It can be immediately seen that more than one sector is macroscopically occupied across the whole doping range. The peak, therefore, is formed and persists in a region of mixed valence for the $f$-orbitals. 

The Boltzmann weight analysis also offers an intuitive explanation for the evolution of the spectral features in the doping range $\nu \in [1,2]$. First, as reasonably expected, it shows that the most occupied sector is $N=5$. This corresponds to the situation where the $f$-subspace has 1 electron more than half-filling, which is consistent with the overall doping $\nu > 1$.
The red (dotted) and green (dashed) lines in \prettyref{sifig:compressibility}(left) highlight two interesting doping values: first, at around $\nu=1.4$, the statistical weight of sector $N=6$ overtakes that of $N=4$. These two sectors are respectively at doping $2$ and  half-filling in the $f$-space. On the spectral density plot, this corresponds to the doping at which the spectral weight undergoes a significant reorganization: at negative frequency, the left shoulder peak fades away. This feature (cfr. Fig.~1 in the main text) is the remnant of the zero-bias resonance in the doping range $[0,1]$, associated to the dominant $N=4$ sector, which is progressively shifted to negative frequencies by increasing occupation. At the same time, the right shoulder peak, seen in the left panel of \prettyref{sifig:compressibility} and highlighted by a green line in  Fig.~1, merges with the zero-bias resonance.
A second relevant spectral reorganization occurs for $\nu=1.87$: here, the central peak is depinned from $\omega=0$ and pushed towards negative frequency, while a new right shoulder peak starts to emerge at around 10 meV. This corresponds to a switch in the dominant weight sector, whereby $N=6$ overtakes $N=5$, and marks the beginning of the cascade behavior around $\nu=2$.

\subsection{Origin of the ``outer'' gap}

\begin{figure}
    \centering
    \includegraphics{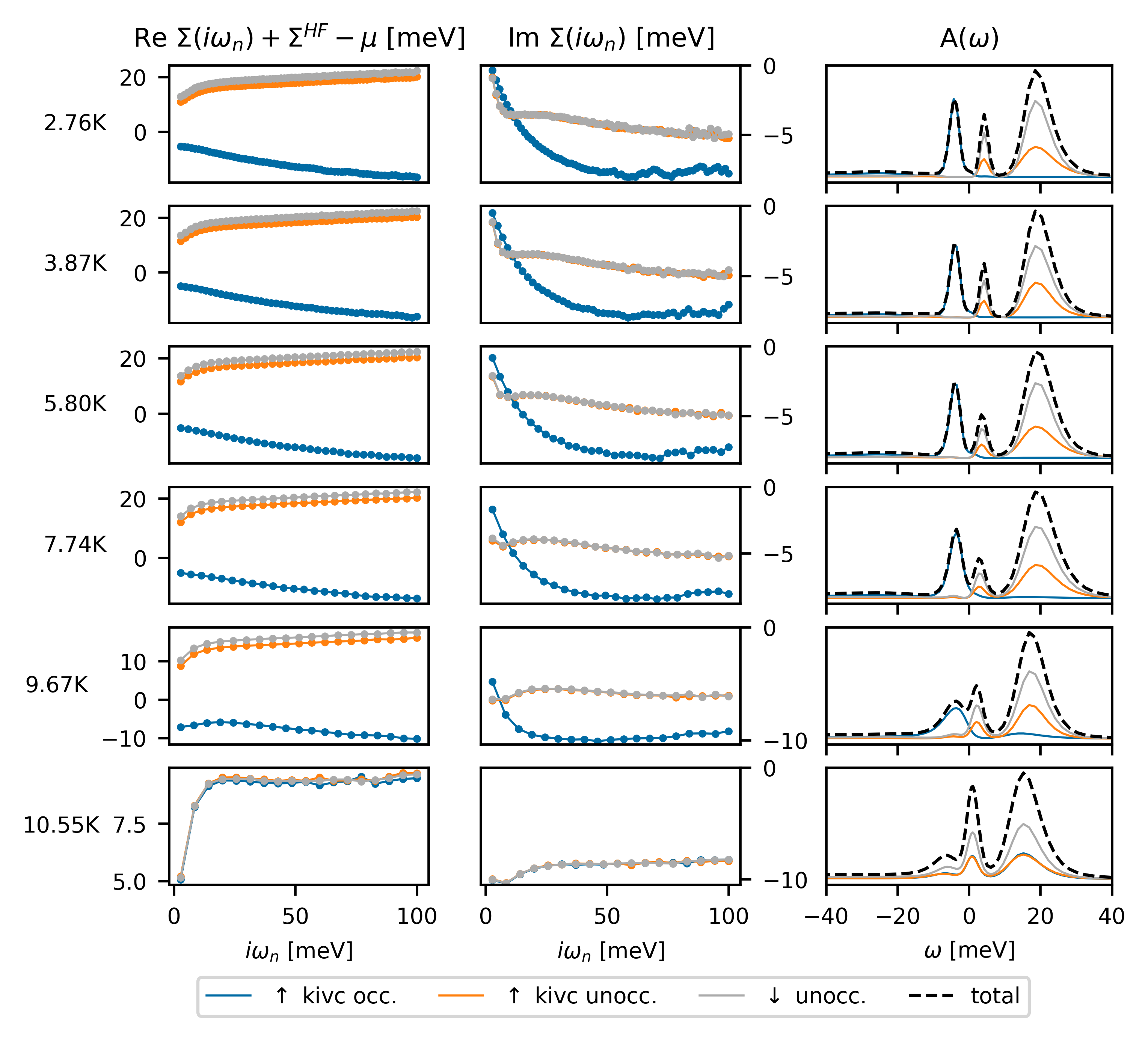}
    \caption{The real (left panel) and imaginary (middle panel) part of the Matsubara self-energies and the analytically-continued spectral functions (right panel) at $\nu=-2.2$ at temperatures ranging from $\approx{3}~K$ to $\approx{11}~K$ in the K-IVC phase. The down spin sector (gray) is preferentially unoccupied, while in the up spin sector one K-IVC sector is preferentially occupied (blue) while the other (orange) is preferentially unoccupied. At $10.55$~K, the system is no longer ordered.}
    \label{sifig:self-energies-and-gully}
\end{figure}

The experiment finds that the outer gap appears at temperatures below about $5$~K and in a one-to-one correspondence with Kekul\'{e} distortion. Since Kekul\'{e} distortion is a feature of the inter-valley coherent incommensurate Kekul\'{e} spiral (IKS) state, this suggests that the outer gap feature is related to inter-valley coherence. Furthermore, the electronic spectra obtained from DMFT simulations in the inter-valley coherent state show a gap with a qualitatively similar filling-dependence as the outer gap seen in the experiment (compare \prettyref{fig: fig2}a and 2c), further supporting that these gaps seen in the STM and DMFT spectra are of the same origin---inter-valley coherent order. 
In \prettyref{sifig:self-energies-and-gully}, we analyze the Matsubara self-energies and the orbital-resolved spectral functions in the K-IVC phase to understand the outer gap. The role of ordering as the origin of the gap is immediately apparent from the orbitally-resolved spectral functions, where we see that the spectral weight on the left (right) side of the gap is made up of the preferentially occupied (unoccupied) spin-valley flavors under the chosen inter-valley coherence order. The gap opening is therefore driven by the real part of the self-energy, where we see that the high-frequency constant (Hartree) term differs by $\approx{20}$~meV in the occupied and unoccupied flavors. 

\begin{figure}
    \centering
    \includegraphics[width=\textwidth]{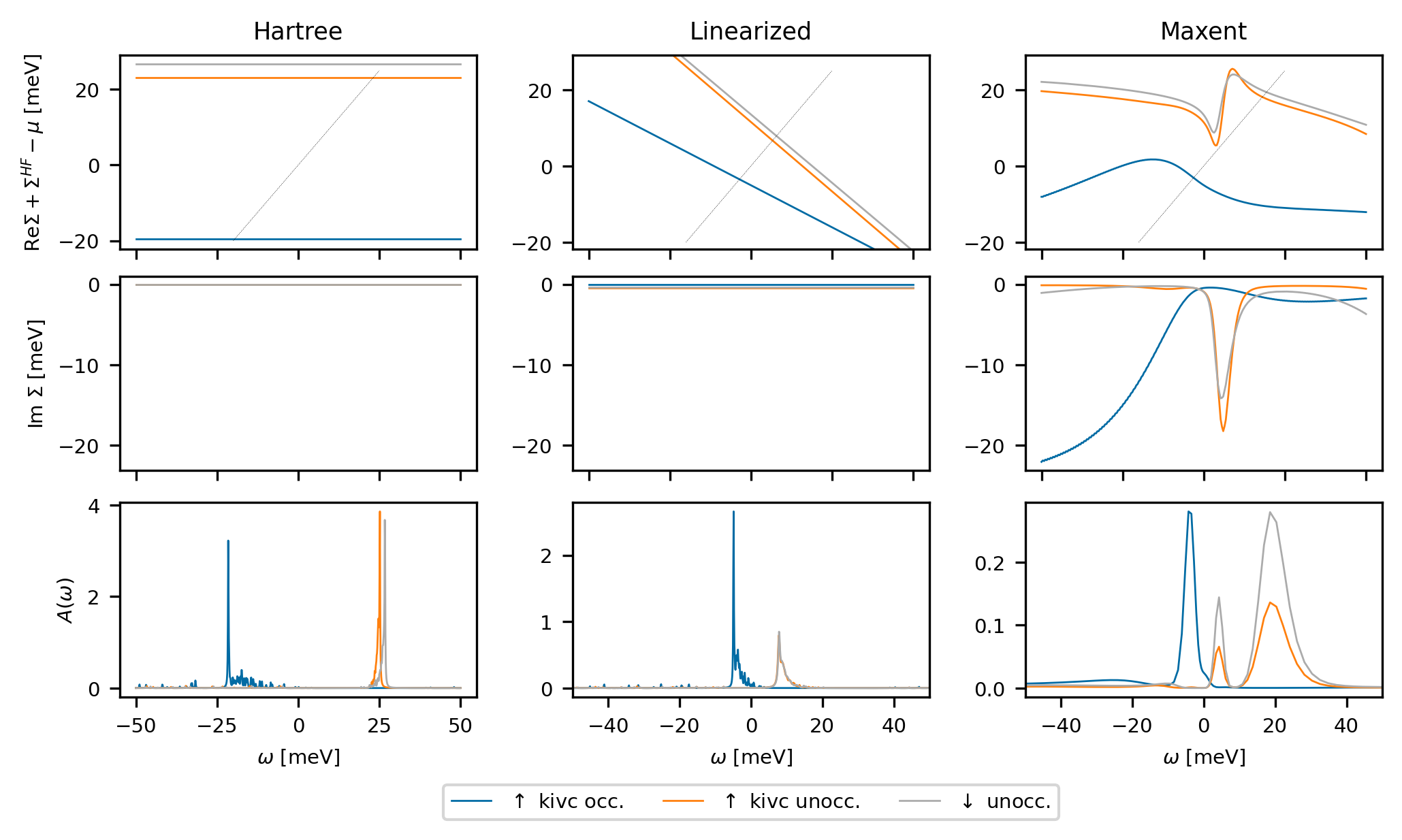}
    \caption{The real (top panel) and imaginary (middle panel) parts of the real-axis self-energies and the corresponding spectral functions (bottom panel) for three different levels of approximation: Hartree-Fock (left), the linearized ansatz (center) and analytical continuation with the maximum entropy method (right). Intersections of the real part of the self-energy with the dotted line of slope unity mark the locations of the poles of the Green function.}
    \label{sifig:origin-of-gap}
\end{figure}

However, ordering is not sufficient to capture all the features of the gap. First, the gap is much smaller than $20$~meV. Second, the gap is symmetric around zero. Both of these are the result of dynamical correlations, captured in DMFT by the frequency dependence of the self-energy, and finite-lifetime effects, captured in the imaginary part of the real-axis self-energy. In order to disentangle the effects of ordering and correlations in the opening of the gap, we extract the spectral function at $\nu=-2.2$ and $T=3.87$~K at three degrees of approximation in \prettyref{sifig:origin-of-gap} that we call Hartree-Fock, Linearized, and Maxent. Following~\citeSM{rai2024-for-si}, we transform the $f$-electron basis to the {\it natural basis} such that the self-energy in the new basis is diagonal chosen symmetry-broken state. Using $\Sigma_\xi$ to label the diagonal components of the self-energy in this basis, the three approximations are defined as:
\begin{align}
    \textrm{Hartree-Fock}:& \hspace{3em}\Sigma_\xi(\omega) \to \textrm{Re}\Sigma_\xi(i\infty)\\
    \textrm{Linearized}:& \hspace{3em}\Sigma_\xi(\omega) \to \Sigma_\xi(i\omega_0) + (1-Z_\xi^{-1})\omega \\
    \textrm{Maxent}:&  \hspace{3em}\Sigma_\xi(\omega) \to \textrm{maxent}\left[\Sigma_\xi(i\omega_n)\right] ,
\end{align}
where $Z_\xi = 1-\left.\frac{\partial \textrm{Im}\Sigma_\xi(i\omega_n)}{\partial i\omega_n}\right|_{i_{\omega_n}\to0}$ is the quasiparticle weight, and maxent[$\Sigma_\xi(i\omega_n)$] refers to the full analytically continued self-energy on the real axis using the maximum entropy method. Despite the fact that Green functions (and self-energies) are analytic on the complex plane, numerically reconstructing them on the real axis from data on the Matsubara axis involves an ill-posed matrix inversion. The maximum entropy method regularises this ill-posed matrix inversion by casting it as the minimisation of a cost function supplemented by a suitably chosen entropic term. We use the TRIQS implementation~\cite{kraberger2017maximum}.

In the left panel of \prettyref{sifig:origin-of-gap}, we plot the self-energies and spectral function within the Hartree-Fock approximation. I.e. we include ordering effects but no correlations and no renormalization. Such a static self-energy captures the opening of the gap, since the occupied and unoccupied flavors are pushed down or up in energy by the real part of the self-energy. However, the gap is far too large, and the transfer of unoccupied spectral weight to the Hubbard peak is missing. 

In the middle panel of \prettyref{sifig:origin-of-gap}, we apply the linearized ansatz. This approximation includes lifetime effects through $\textrm{Im}\Sigma_\xi(i0)$ and expands the real part of the real-axis self-energy to linear order around zero, with the slope given by $(1 - Z_\xi^{-1})$. This approximation, which loses validity at higher frequencies, captures renormalization effects around zero, encoded in the quasiparticle weight $Z$. In this case, $Z$ squeezes the peaks corresponding to the occupied and unoccupied sectors, reducing the size of the gap. Note that in the absence of spin-valley ordering, the Fermi-liquid--like self-energy would result in a quasiparticle peak at the Fermi level with weight $Z$. Therefore, the approximate particle-hole symmetry of the gap (pinning of the gap center to zero) is of the same origin as the pinning of the quasiparticle peak to zero in the paramagnetic phase. In other words, the gap opens within the quasiparticle peak of the intermediate valence state in the paramagnetic phase.

The linearized self-energy only allows one pole per $k$ point and cannot capture the transfer of spectral weight of the unoccupied sectors to the Hubbard peak. The full analytically continued self-energies using the maximum entropy method are in the right panel of \prettyref{sifig:origin-of-gap}. Here, we also capture the depletion of spectral weight due to Mott-Hubbard splitting with an upper Hubbard peak around 20~meV. Due to the intermediate valence nature of the state, the lower Hubbard peak and the quasiparticle peak conflate around zero. 

\bibliographystyleSM{naturemag}
\bibliographySM{SI-theory}

\end{document}